\documentclass[12pt,preprint]{aastex}

\begin{document}

\newcommand{\OI}{~1}
\newcommand{\NaIa}{~2}
\newcommand{\NaIb}{~3}
\newcommand{\MgI}{~4}
\newcommand{\AlI}{~5}
\newcommand{\SiIGa}{~6}
\newcommand{\SiITh}{~4}
\newcommand{\CaI}{~7}
\newcommand{\TiIa}{~8}
\newcommand{\TiIb}{~9}
\newcommand{\TiIc}{10}
\newcommand{\TiII}{11}
\newcommand{\VI}{12}
\newcommand{\CrI}{11}
\newcommand{\FeIb}{13}
\newcommand{\FeIc}{14}
\newcommand{\FeId}{15}
\newcommand{\FeIe}{16}
\newcommand{\FeIOB}{17}
\newcommand{\FeIBKa}{18}
\newcommand{\FeIBKb}{19}
\newcommand{\FeII}{20}
\newcommand{\CoI}{21}
\newcommand{\NiI}{22}
\newcommand{\BaII}{23}
\newcommand{\LaII}{24}
\newcommand{\EuII}{25}

\title{CHEMICAL ABUNDANCES OF THREE METAL-POOR GLOBULAR CLUSTERS
(NGC~6287, NGC~6293, AND NGC~6541) IN THE INNER HALO}

\author{Jae-Woo Lee\altaffilmark{1,2,3,4}
\and Bruce W. Carney\altaffilmark{1,4}}

\altaffiltext{1}{Department of Physics \& Astronomy,
University of North Carolina, Chapel Hill, NC 27599-3255; 
jaewoo@astro.unc.edu; bruce@physics.unc.edu}
\altaffiltext{2}{Center for Space Astrophysics, 
Yonsei University, Shinchon-dong 134, Seoul 120-749, Korea}
\altaffiltext{3}{PMA Division, California Institute of Technology, 
Mail Stop 405-47, Pasadena, CA 91125, USA; jaewoo@srl.caltech.edu}
\altaffiltext{4}{Visiting Astronomer, Cerro Tololo Inter-American 
Observatory, National Optical Astronomy Observatories,
which are operated by the Association of Universities for Research 
in Astronomy, Inc., under contract with the National Science Foundation.}

\begin{abstract}
We present a chemical abundance study of three inner old halo clusters 
NGC~6287, NGC~6293, and NGC~6541, finding
[Fe/H] = $-$2.01 $\pm$ 0.05, $-$1.99 $\pm$ 0.02, and $-$1.76 $\pm$ 0.02
(internal), respectively, and our metallicity measurements are
in good agreement with previous estimates.
We also present the radial velocity measurements of the clusters.
Our radial velocity measurements for NGC~6293 and
NGC~6541 are in good agreement with previous measurements,
however, our radial velocity measurement for NGC~6287
is almost 80 km~s$^{-1}$ larger than the previous measurement.

The mean $\alpha$-element abundances of our program clusters are
in good agreement with other globular clusters, confirming previous results.
However, the individual $\alpha$-elements appear to follow different trends.
The silicon abundances of the inner halo clusters appear to be enhanced and
the titanium abundances appear to be depleted compared to the intermediate
halo clusters. Our results also appear to oppose to those of metal-rich bulge
giants studied by McWilliam and Rich, who found that bulge giants
are titanium enhanced and silicon deficient.
In particular, [Si/Ti] ratios appear to be related to Galactocentric distances,
in the sense that [Si/Ti] ratios decrease with Galactocentric distance. 
We propose that contributions from different masses of the SNe~II progenitors
that enriched proto-globular cluster clouds' elemental abundances and
the different initial physical environments surrounding 
the proto-globular clusters clouds
are responsible for this gradient in [Si/Ti] ratios versus 
Galactocentric distances of the ``old halo" globular clusters. 
On the other hand, our program clusters' enhanced $s$-process
elemental abundances suggest that the formation timescale of
our program clusters might be as short as a few times 10$^8$ yr
after the star formation is initiated in the Galaxy's central regions,
if the $s$-process site is intermediate mass AGB stars.

\end{abstract}


\section{INTRODUCTION}
Detailed elemental abundance studies of globular clusters may
provide strong constraints on the Galaxy formation picture. For
example, a metallicity gradient would imply that the Galaxy formed
via a slow dissipational process. A constant and enhanced
[$\alpha$/Fe] versus [Fe/H] relation may indicate that the
globular clusters must have formed simultaneously within a couple
of gigayears (Wyse \& Gilmore 1988; Wheeler et al.\ 1989; Carney 1996), 
so that their proto-globular cluster clouds (PGCCs) were not contaminated
by SNe~Ia products.
The abundance ratio of $r$-process elements to $s$-process elements,
such as [Ba/Eu] and [La/Eu], as a function of metallicity in
globular cluster systems, may also suggest how rapidly they were
polluted by the low-\ or intermediate-mass stars before they formed.

In spite of the importance of the chemical abundance studies, 
few high resolution spectroscopic studies of the globular clusters 
near the Galactic center have been performed due to 
the observational limitations set by large interstellar reddening.
Geisler (1988) studied NGC~6541 employing the Washington
photometric system and obtained [Fe/H] = $-$0.99 $\pm$ 0.20. The
high metallicity for NGC~6541 led Geisler (1988) to claim that
NGC~6541 is the one of the examples of the second parameter
problem since it has a BHB morphology for its metallicity.
Rutledge et al. (1997a) studied NGC~6541 using the Ca~II triplet
with low resolution spectra and obtained [Fe/H] = $-$1.79 $\pm$
0.02 on the Zinn \& West (1984) metallicity scale. 
In this case, a second parameter is unnecessary to explain
the HB morphology of NGC~6541.
Kennedy, Bates, \& Kemp (1998) employed high resolution
spectroscopy for several stars in NGC~6541, but their study
focused on the interstellar medium towards NGC~6541 using
spectral regions near Na~I D lines.

In this paper, we explore the detailed elemental abundances for
RGB stars in the metal-poor inner halo globular clusters 
NGC~6287 ($R_{GC}$ = 1.6 kpc), NGC~6293 ($R_{GC}$ = 1.4 kpc), 
and NGC~6541 ($R_{GC}$ = 2.2 kpc) using high resolution spectra. 
Lee et al.\ (2001) and Lee, Carney, \& Heasley (2002, in preparation)
employed HST NIC3 and PC2 photometry of these clusters to show
that they essentially have the same age as the one of the oldest
globular clusters in our Galaxy, M92. A comparative elemental
abundance study between these three clusters and the intermediate
halo clusters (or the metal-poor halo stars) will provide clues of
the early chemical enrichment history of the inner part of our
Galaxy.

\section{OBSERVATIONS AND DATA REDUCTION}
We selected our program RGB stars from Stetson \& West (1994),
Janes \& Heasley (1991), and Alcaino (1979) for NGC~6287, NGC~6293,
and NGC~6541, respectively.
We obtained high S/N ($\geq$ 95) echelle spectra using
the CTIO 4-meter telescope and its Cassegrain echelle spectrograph.
The Tek 2048 $\times$ 2048 CCD, 31.6~lines/mm echelle grating,
long red camera, and G181 cross-disperser were employed
for our observations.
The slit width was 150 $\mu$m, or about 1.0 arcsec, that projected to
2.0 pixels and which yielded an effective resolving power $R$ = 28,000.
Each spectrum had complete spectral coverage from 5500 to 7850~\AA\ for
the 1998 run and 5700 to 8000~\AA\ for the 1999 run.
All program star observations were accompanied by flat lamp,
Th-Ar lamp, and bias frames.
We also obtained spectra of rapidly rotating early type stars
in order to remove telluric absorption features.
During the 1998 run, the seeing conditions were extremely poor,
no better than 2-2.5 arcsec for 5 nights.
The seeing conditions were slightly better, with the mean of 1.5 arcsec,
during the 1999 run.
The basic photometric data for the program RGB stars and
the journal of observations are given in Table~1.

The raw data frames were trimmed, bias-corrected, and flat-fielded
using the IRAF\footnote{IRAF (Image Reduction and Analysis
Facility) is distributed by the National Optical Astronomy
Observatory, which is operated by the Association of Universities
for Research in Astronomy, Inc., under contract with the National
Science Foundation.} ARED and CCDRED packages. The scattered light
was also subtracted using the APSCATTER task in ECHELLE. The
echelle apertures were then extracted to form 1-d spectra, which
were continuum-fitted and normalized, and a wavelength solution
was applied following the standard IRAF echelle reduction
routines. The telluric line removal was performed by dividing a
program star spectrum by that of rapidly rotating early type
stars.

Equivalent widths were measured mainly by the direct integration of
each line profile using the SPLOT task in IRAF ECHELLE package.
The equivalent widths for our program stars are listed in
Table~2.

\section{ANALYSIS}
In our elemental abundance analysis, we use the usual spectroscopic
notations that
$[A/B] \equiv \log (N_A/N_B)_{star} - \log (N_A/N_B)_{\sun}$,
and that $\log n(A) \equiv \log (N_A/N_H) + 12.00$ for each element.
For the absolute solar iron abundance, we adopt $\log n$(Fe) = 7.52
following the discussion by Sneden et al.\ (1991).

\subsection{Line Selection and Oscillator Strengths}
For our line selection, laboratory oscillator strengths
were adopted whenever possible,
with supplemental solar oscillator strength values.
In addition to oscillator strengths, taking into account
the damping broadening due to the van der Waals force, we adopted
the Uns\"old approximation with no enhancement.
We list the source of oscillator strengths in Table~3.

The abundance analysis mainly depends on the reliability
of the oscillator strength values for the Fe~I and Fe~II lines,
since not only the metallicity scale but also the
stellar parameters, such as spectroscopic temperature, surface
gravity, and microturbulent velocity, will be determined by using
these lines. We mainly relied upon the extensive laboratory
oscillator strength measurements by the Oxford group 
(Blackwell et al.\ 1979; 1982b, 1982c, 1986a).
We also used oscillator strength values measured by O'Brian et al.\ (1991)
and the Hannover group (Bard, Kock, \& Kock 1991; Bard \& Kock 1994).
In our iron abundance analysis, we consider the Oxford group's measurements
(the absorption method) as primary oscillator strengths and
oscillator strength measurements that relied on emission methods
(O'Brian et al.\ 1991; Bard, Kock, \& Kock 1991; Bard \& Kock 1994)
as supplemental. Therefore, the oscillator strengths by O'Brian et al.\
and the Hannover group were scaled with respect to those by
the Oxford group as a function of excitation potential
(de Almeida 2000, private communication),
\begin{eqnarray}
\log gf &=& \log gf(OB) - 0.017, \nonumber \\
\log gf &=& \log gf(H91) - 0.015 - 0.009 \chi, \nonumber \\
\log gf &=& \log gf(H94) - 0.027 - 0.009 \chi,
\label{eqn6:gfscale}
\end{eqnarray}
where the excitation potential $\chi$ is given in electron volts.
Blackwell, Smith, \& Lynas-Gray (1995) also pointed out that there
appears to exist a slight gradient in the excitation potential
between oscillator strengths by the Oxford group and those by the
Hannover group, with $\log gf(Oxford) = \log gf(Hannover) - 0.021
- 0.006\chi$.

For neutral titanium lines, we relied on the laboratory
measurements by the Oxford group (Blackwell et al.\ 1982a, 1983,
1986b). It should be noted that the original Oxford $gf$-values
have been increased by +0.056~dex following Grevesse, Blackwell,
\& Petford (1989). They discussed that their original $gf$-values
relied on the inaccurate lifetime measurements and the absolute
$gf$-values should be revised based on the new measurements.

Contrary to other elements possessing HFS components,
HFS components should be considered in the barium abundance analysis
because Ba~II lines are usually very strong even in metal-poor stars and
the desaturation effects due to HFS components become evident
(see for example, McWilliam 1998).
We adopted the Ba~II HFS components and
oscillator strengths of Sneden et al.\ (1997).
Note that Sneden et al.\ (1997) did not publish their HFS line list but
the electronic version of Ba~II HFS components was kindly provided by
Chris Sneden and Inese Ivans (2000, private communication).

\subsection{Stellar Parameters and Model Atmospheres}
Having good stellar parameters, such as the effective temperature and
the surface gravity, is very important in any stellar abundance study,
since the absolute or the relative elemental abundance scale will depend
on the input stellar parameters.
For our analysis, we rely on spectroscopic temperatures and
photometric surface gravities.

The initial estimates of the temperature of program stars were estimated
using $BV$ photometry of our program clusters (Stetson \& West 1994,
Janes \& Heasley 1991, and Alcaino 1979,
for NGC~6287, NGC~6293, and NGC~6541, respectively) and
the empirical color-temperature relation given by
Alonso, Arribas, \& Martinez-Roger (1999).
Since their relation depends slightly on the metallicity, 
we adopted [Fe/H] = $-$2.05, $-$1.90, and $-$1.80 for NGC~6287, 
NGC~6293, and NGC~6541 (Harris 1996).
To estimate the dereddened color, we adopt $E(B-V)$ = 0.62, 0.40, 
and 0.14 for NGC~6287 (Lee et al.\ 2001), NGC~6293, and NGC~6541
(Lee, Carney, \& Heasley 2002, in preparation), respectively (see also Table~4).
It should be noted that our $E(B-V)$ values are estimated
using HST NIC3 photometry (NGC~6287) and PC2 photometry 
(NGC~6293 and NGC~6541) are in good agreement with 
those of Harris (1996) to within 0.02 mag.
Our program clusters have large interstellar reddening values because
they are located near the Galactic center and they even suffer from
the differential reddening.\footnote{Lee et al.\ (2001) discussed
differential reddening in NGC~6287. 
Based on the mean turn-off colors in two different pointings of 
HST NIC3 observations, Lee et al.\ suggested that 
$\Delta E(B-V)$ $\approx$ 0.07 -- 0.09 mag exists across NGC~6287, 
confirming the previous result of Stetson \& West (1994).}
Therefore, the uncertainty in the interstellar reddening 
is the major source of error in our photometric temperature estimation.
Generally, an uncertainty of 0.1 mag in $E(B-V)$ results in 
an uncertainty of $\approx$ 100 -- 130 K in photometric effective
temperature.

To derive photometric surface gravity in relation to that of Sun, 
we use $\log g_{\sun}$ = 4.44 in cgs units,
$M_{bol,\sun}$ = 4.74 mag, and $T_{eff,\sun}$ = 5777 K for the Sun
(Livingston 1999) and we assume the stellar mass $M$ = 0.8 $M_\sun$.
We use the empirical relation given by Alonso et al.\ (1999)
to estimate the bolometric correction.
We adopt $(m-M)_0$ = 14.35, 14.61, and 14.19 mag for NGC~6287 
(Lee et al.\ 2001), NGC~6293, and NGC~6541 
(Lee, Carney, \& Heasley 2002, in preparation).
Our $(m-M)_0$ values are 0.30, 0.11, and 0.05 mag smaller than those 
of Harris for NGC~6287, NGC~6293, and NGC~6541, respectively.
In general, an uncertainty of 0.3 mag in $(m-M)_0$ results in
an uncertainty of 0.1 dex in $\log g$, in the sense that a short
distance scale results in a high surface gravity.

As an independent test, we compared the spectroscopic temperature
and surface gravity of M92 RGB stars by Sneden et al.\ (1991) to
those calculated using the photometric method above.
Using M92 RGB stars provides an advantage that the interstellar reddening
towards M92 is negligibly small, $E(B-V)$ = 0.02.
Thus, the uncertainty raised by the interstellar reddening
in the photometric temperature and surface gravity estimates
will be minimized. Our calculations showed that
the photometric and spectroscopic temperature scales
are in very good agreement with
$\Delta$[$T_{eff}$(spectroscopic) $-$ $T_{eff}$(photometric)] = 37 $\pm$ 11 K.
We calculate the photometric surface gravity for
0.80 $M_\odot$ and 0.85 $M_\odot$,
which are roughly the stellar evolutionary mass for the RGB stars in
the oldest metal-poor clusters (Bergbusch \& VandenBerg 2001).
Both cases are in very good agreement with the surface
gravity measurements by Sneden et al.\ (1991).
For our photometric surface gravity estimates,
a stellar mass of 0.80 $M_\odot$ is assumed for each program star.

With initial photometric temperature and surface gravity estimates,
72-depth plane-parallel LTE model atmospheres were computed
using the program ATLAS9, written and supplied by Dr.\ R.\ L.\ Kurucz.
Assuming that the star would prove to be metal-poor,
the model atmospheres were computed using opacity distribution functions
and abundances with enhanced abundances of all the ``$\alpha$" elements
(O, Ne, Mg, Si, S, Ar, Ca, and Ti) by 0.4~dex.
The ``$\alpha$" element enhancements are important since
several of these elements are quite abundant and are major
electron donors to the H$^{-}$ opacity.
During our model computation, the convective overshoot was turned on.
This provided our initial model atmospheres for each program star.

The abundance analysis was performed using the programs WIDTH9
(written and supplied by Dr.\ R.\ L.\ Kurucz) and MOOG (Sneden 1973).
Adopting the photometric temperature and surface gravity 
as our initial values, we began by restricting the analysis to 
those Fe~I lines with log(W$_{\lambda}$/$\lambda$) $\leq$ $-5.2$
({\em i.e.} for  the linear part of the curve of growth),
and comparing the abundances as a function of excitation potential.
New model atmospheres were computed with a slightly different effective
temperature until the slope of the log~n(Fe~I)  versus $\chi$ relation
was zero. The stronger Fe~I lines were then added and the microturbulent 
velocity $v_{turb}$ altered until the $\log$ n(Fe~I) versus 
$\log$(W$_{\lambda}$/$\lambda$) relation had zero slope.
Finally, we analyzed the gravity-sensitive Fe~II lines,
recomputing new model atmospheres by altering the surface gravity
until the iron abundances derived from the Fe~I lines agreed with
those derived from the Fe~II lines.

Table~4 shows comparisons of temperature and surface gravity between 
the photometric (Alonso et al.\ 1999) and the spectroscopic methods 
in our program stars. The temperatures agree well between the two methods,
with the exception of NGC~6293-2673. 
The discrepancy in temperature appears to be related to the interstellar
reddening value of the clusters, suggesting that the interstellar reddening
values may be slightly incorrect or there exists a differential reddening
effect. (It is also likely due in part to the quality of spectra. Our spectra
for the NGC~6541 RGB stars are superior to our spectra for the other clusters.)
Figure~1 shows the comparisons of our spectroscopic
$T_{eff}$ versus photometric and spectroscopic $\log g$ results
to those of model isochrones\footnote{In the Figure,
we adopted the model isochrones for 14 Gyr. The age does not significantly
affect the RGB loci of the model isochrones with $>$ 10 Gyr.}
for [Fe/H] =$-$2.14 and $-$1.84, and [$\alpha$/Fe] = +0.3
(Bergbusch \& VandenBerg 2001).
In the Figure, we also show $T_{eff}$ and $\log g$ of M92 RGB stars
(Sneden et al.\ 1991). The mean metallicity of the M92 RGB stars is
[Fe/H] $\approx$ $-$2.3 (Sneden et al.\ 1991, 2000b)
and is more metal-poor than our three program clusters.
Our photometric $\log g$ results agree well with those of model isochrones,
while our spectroscopic $\log g$ results appear to be too small.
The disagreement in the surface gravity is rather large and
it appears to be related to NLTE effects as suggested by others
(see, for example, Nissen et al.\ 1997; Allendo Prieto et al.\ 1999).
Since metal-poor stars have much weaker metal-absorption in the ultraviolet,
more non-local UV flux can penetrate from the deeper layers.
This flux is vital in determining the ionization equilibrium of the atoms,
resulting in deviations from LTE.
Nissen et al.\ (1997) claimed that surface gravities of metal-poor 
dwarfs and subgiants derived from the spectroscopic method, which demands
that Fe~I and Fe~II lines should provide the same iron abundance,
are a factor of two or three ($\Delta \log g \approx$ 0.3 -- 0.5)
smaller than those from the Hipparcos parallaxes.
Allendo Prieto et al.\ (1999) also claimed that spectroscopic gravities and
those from the Hipparcos parallaxes are in good agreement for stars
in the metallicity range $-1.0 <$ [Fe/H] $<$ +0.3, while
large discrepancies can be found for stars with metallicities below
[Fe/H] = $-$1.0, in the sense that the spectroscopic method provides
lower surface gravities. Therefore, we rely on photometric gravities 
for our abundance analysis. It should be noted that the photometric surface
gravity estimate is rather insensitive to the uncertainties 
in interstellar reddening values. 
In general, an uncertainty of 0.1 mag in $E(B-V)$ results in
an uncertainty of 0.1 dex in the derived photometric surface gravity,
in the sense that the photometric surface gravity increases with
decreasing $E(B-V)$.

The program MOOG provides a powerful means of synthetic spectrum fitting,
which is necessary for the HFS analysis in particular for
Ba~II lines, for example via the BLENDS driver.
To see if there is any systematic difference between the results from
WIDTH9 and those from MOOG, we compared the Fe~I abundances
for NGC~6287-1491 using the same model atmosphere and atomic data
and we obtained consistent results between the two, with
$\Delta$[$\log$~n(Fe~I)$_{\mathrm{WIDTH9}}$ $-$
$\log$~n(Fe~I)$_{\mathrm{MOOG}}$] = 0.00 $\pm$ 0.01.
Therefore, we consider that WIDTH9 and MOOG give consistent
results to within 0.01 dex and we do not differentiate between
the two in the following discussions.

\section{RESULTS}
\subsection{Radial Velocity Measurements}
Although radial velocity measurements were not our primary goal,
they provide a very useful criterion for cluster membership
of the program stars and allow us to update the somewhat uncertain
radial velocities of our program clusters.
The radial velocity is of particular importance
in the kinematic study of inner halo globular clusters.
Since there are not yet any proper motion studies of inner halo clusters,
the kinematic properties of these clusters have been investigated
via statistical analyses using their radial velocities
(see, for example, Harris 2001).
Previous studies of the radial velocities have been based
on low resolution spectra,
where the internal measurement error is much larger than ours.

In order to determine the radial velocity, we cross-correlated about
20 orders in our program star spectra with the same orders observed
each night for the twilight sky.  We show heliocentric radial velocities of
individual stars in Table~5.
As can be seen in the Table, NGC~6541~I-21 is not a member star of
the cluster and we therefore exclude it from our mean velocity calculations.
Our mean heliocentric radial velocities for NGC~6287, NGC~6293,
and NGC~6541 are $-$288.8 $\pm$ 3.5 km~s$^{-1}$,
$-$151.9 $\pm$ 6.7 km~s$^{-1}$, and $-$167.5 $\pm$ 5.0 km~s$^{-1}$,
respectively.  (The errors are those of mean.)
It should be noted that large uncertainties in our velocity measurements
are mainly due to the cluster internal velocity dispersions.

Hesser, Shawl, \& Mayer (1986) studied the radial velocities of
these three clusters using low resolution image-tube spectrograms,
and they obtained $-$208 $\pm$ 16 km~s$^{-1}$, $-$143 $\pm$ 17 km~s$^{-1}$,
$-$158 $\pm$ 7 km~s$^{-1}$ for NGC~6287, NGC~6293, and NGC~6541,
respectively. Their radial velocities for NGC~6293 and NGC~6541 are
in agreement to within the measurement errors.
The discrepancy in the radial velocity of NGC~6287, however, is so large
that it can not be easily explained by measurement errors.
During the 1998 observing run, we obtained a spectrum of HD~166161,
whose radial velocity is well known. Using the same procedure
described above, we derived  $v_r$ = 68.2 $\pm$ 0.7 km~s$^{-1}$
for HD~166161 and this is in excellent agreement with the previous
measurements by others (68.4 $\pm$ 0.3 km~s$^{-1}$, Bond 1980).
Therefore, we conclude that our radial velocity measurement for NGC~6287
is correct and differs by almost 80 km~s$^{-1}$ larger than
that of Hesser et al.\ (1986).

Rutledge et al.\ (1997b) measured the radial velocities for 11 stars in
NGC~6541 and obtained a mean radial velocity for the cluster of
$v_r$ = $-$163.5 $\pm$ 12.4 km~s$^{-1}$.
Their radial velocity for NGC~6541 is in good agreement with our value.

\subsection{Elemental Abundances and Error Analysis}
In Tables~6, 7 and 8, we present the elemental abundances of our program
clusters. In the Tables, the elemental abundances using the photometric
surface gravities (column $\log g_P$) and the spectroscopic gravities
(column $\log g_S$), the uncertainty, and the number of absorption lines
used for each element are listed.
In both cases, we adopt the spectroscopic temperatures given in Table~4.
The [el/Fe] ratios for neutral elements are estimated from [el/H]
and [Fe~I/H] ratios, with the exception of oxygen.
The [el/Fe] for singly ionized elements (Ti~II, Ba~II, La~II, and Eu~II)
and oxygen are estimated from [el/H] and [Fe~II/H] ratios
(see, for example, Ivans et al.\ 2001).
The uncertainty quoted is for a single line and, therefore,
that of each element is given by $\sigma/\sqrt{n}$,
where $\sigma$ is the uncertainty per line and
$n$ is the number of absorption lines used for each element.
The mean iron abundance [Fe/H]$_{avg}$ is defined to be
the unweighted average of [Fe/H]$_\mathrm{I}$ and [Fe/H]$_\mathrm{II}$.
In Table~9, we show the mean results per cluster.
It should be noted that we rely on the abundances using
the photometric surface gravity and we adopt Fe~II abundances
for the iron abundances of our program stars since
the Fe~II abundance is thought to be less sensitive to NLTE conditions
(Th\'evenin \& Idiart 1999).

In Table~10, we show estimated errors resulting from uncertainties
in the input model atmosphere $\delta T_{eff}$ = $\pm$100 K, 
$\delta \log g$ = $\pm$0.3 and $\delta v_{turb}$ = $\pm$0.3 km s$^{-1}$, 
which are appropriate for our analysis.
In the fourth column of the Table, we also show abundance errors
by shifting $\delta T_{eff}$ = $\pm$100 K. In this case, however,
we also alter the photometric $\log g$ and $v_{turb}$ to meet criteria 
that we used in the previous section [photometric surface gravity and 
the zero slope in the $\log$ n(Fe~I) versus 
$\log$ (W$_{\lambda}$/$\lambda)$ relation].
The errors in the fourth column have similar values as those 
in the first column. It should be noted that O, Si, Ba, La, and Eu are 
very  sensitive to $T_{eff}$ resulting in 
$\mid$$\delta$[el/Fe]/$\delta T_{eff}$(100 K)$\mid$ 
$\approx$ 0.10 -- 0.14 dex.
Also importantly, the [Si~I/Ti~I] ratio is very sensitive to $T_{eff}$.
An uncertainty of 100 K in $T_{eff}$ results in $\approx$ 0.2~dex 
in the [Si~I/Ti~I] ratio.
This is mainly due to high sensitivity in the [Ti~I/H] ratio to $T_{eff}$.
On the other hand, the [Si~I/Ti~II] ratio is less sensitive to $T_{eff}$,
$\mid$$\delta$[el/Fe]/$\delta T_{eff}$(100 K)$\mid$ $\approx$ 0.02 dex.
Since we are forced to rely on very strong lines, the barium abundance is very 
sensitive to the microturbulent velocity. Fortunately, the abundance
of the other s-process element, lanthanum, relies on weaker lines.

\subsection{The $\alpha$-elements}
The measurements of the $\alpha$-element abundances provide us
an opportunity to explore the relative ages of the globular clusters.
The enhanced values of these elements in globular clusters
are interpreted as the domination by SNe~II nucleosynthesis
in the proto-stellar material,
while lower values are interpreted as due to the increasing contribution
of SNe Ia, which are thought to appear 10$^9$ or more years later
(Wyse \& Gilmore 1988; Wheeler et al.\ 1989; Carney 1996).

Carney (1996) suggested that the unweighted mean value of
silicon, calcium, and titanium abundances
[$\langle$Si + Ca + Ti$\rangle$/Fe] is the most representative to
the $\alpha$-element abundance in the globular cluster systems.
Since these three elements are not destroyed or produced during the RGB phase,
they are less sensitive to the evolutionary effects,
such as an internal mixing. We, therefore, define the $\alpha$-element
abundance of the globular clusters to be the unweighted mean value of
silicon, calcium, and titanium abundances in the following discussions.

In Figure~2, we show [Si/Fe], [Ca/Fe], [Ti/Fe], and
[$\alpha$/Fe] of our program clusters along with
the ``halo" and the ``disk" globular clusters\footnote{
Zinn (1985) subdivided the globular clusters into the ``halo"
and the ``thick disk" groups at [Fe/H] = $-$0.8.
The halo clusters have an essentially spherical distribution about
the Galactic center and they constitute a pressure supported system 
(a small rotational velocity and a larger velocity dispersion),
while the thick disk clusters have a highly flattened spatial distribution 
and constitute a rotational supported system 
(a larger rotational velocity and a smaller velocity dispersion).
Searle \& Zinn (1978) and Lee, Demarque, \& Zinn (1994)
suggested that the inner halo globular  clusters exhibit 
a tight HB morphology versus [Fe/H] relation,
while the outer halo globular clusters show the second parameter
phenomenon (i.e., a larger scatter in HB type at a given [Fe/H]).
Subsequently, Zinn (1993) subdivided the halo clusters into two groups.
The ``old halo" group obeys the same HB type versus [Fe/H] relationship
as the inner halo clusters while the ``younger halo" group deviates from
this relationship by a significant amount (see Figure~3).
Zinn (1993) and Da Costa \& Armandroff (1995) argued that the old and 
the younger halo groups have different kinematic properties that
the old halo group has a prograde mean rotation velocity with
a smaller velocity dispersion while the younger halo group has a retrograde
mean rotation velocity about the Galactic center with a larger velocity
dispersion. They suggested that the old halo group formed during the collapse
that led ultimately to the formation of the Galactic disk and
the younger halo group were accreted later in time.}
(Gratton 1987; Gratton \& Ortolani 1989;  
Kraft et al.\ 1992, 1995, 1997, 1998; McWilliam et al.\ 1992; 
Sneden et al.\ 1994, 1997, 2000b; Brown et al.\ 1997, 1999;
Ivans et al.\ 1999; Shetrone \& Keane 2000; Habgood 2001)
as a function of [Fe/H]. 
In Table~11, we summarize the mean $\alpha$-element
abundances of our program clusters and other clusters. 
It should be noted that we exclude our program clusters from the mean
values of the ``old halo" globular clusters (OHGCs).
In the Table, the errors are those of the mean. 
The silicon abundances of our program clusters appear
to be slightly more enhanced than those of other clusters.
On the other hand, the titanium abundances of our program clusters 
appear to be slightly depleted.
The titanium abundance of the metal-poor RGB stars
using the neutral titanium lines may suffer from NLTE effects,
such as an over-ionization, and the resultant Ti abundance may be
spurious. However, our Ti abundance analyses using the Ti~II lines
also yield lower titanium abundance scales in our program clusters,
indicating that they are truly titanium deficient.
Note that titanium could be defined partly as an iron-peak
element. Explosive nucleosynthesis calculations of the massive stars
(Woosley \& Weaver 1995) predict that one of the major sources of
the SNe II titanium yield is $^{48}$Cr via
the consecutive electron capture processes.

\subsection{The Heavy Neutron Capture Elements Ba, La, and Eu}
In Figure~4, we show Ba, La, and Eu abundances of
the globular clusters (Gratton et al.\ 1986; Gratton 1987;
Gratton \& Ortolani 1989; McWilliam et al.\ 1992;
Brown et al.\ 1997, 1999; Sneden et al.\ 1997; Kraft et al.\ 1998;
Ivans et al.\ 1999; Shetrone \& Keane 2000; Sneden et al.\ 2000a, 2000b)
and field stars (Burris et al.\ 2000) as a function of metallicity.
In Figure~5, we show [Ba/Eu], and [La/Eu] ratios as a function of [Fe/H]. 
The Figures suggest that the $s$-process elements La and Ba are 
slightly more enhanced in our program clusters than in the field stars.
Figure~6 shows the elemental abundances of Ba, La, and Eu of 
our program clusters and M15 (Sneden et al.\ 2000a).
We also plot the solar abundances of the neutron capture elements with
56 $\leq Z \leq$ 64, those due to the $r$-\ and $s$-process contributions
(Burris 2000). In the Figure, these elemental abundances 
were normalized to the $r$-process element europium. As suggested by others
(Cowan et al.\ 1999; Burris et al.\ 2000; Sneden et al.\ 2000a),
the solar neutron capture element distribution due to the $r$-process
contribution is consistent with those of M15 within the measurement error.
However, the $s$-process elements Ba and La in our program clusters appear
to be slightly enhanced, consistent with Figure~5.
This enhanced $s$-process element distribution of the inner halo globular
clusters may suggest that they might have experienced a different
chemical enrichment history than other halo globular clusters or
the metal-poor field stars.

\section{THE CHEMICAL EVOLUTION OF THE INNER HALO GLOBULAR CLUSTERS}
As Zinn (1993) and Da Costa \& Armandroff (1995) suggested,
if the bulk of ``younger halo" globular clusters (YHGCs)
were accreted into our Galactic halo after the formation of the OHGCs,
elemental abundance patterns in YHGCs may not fully reflect 
the chemical evolution history of our Galactic halo, 
since YHGCs could have experienced very different chemical evolution
history than OHGCs. Also the number of YHGCs studied employing
high-resolution spectroscopy is too small as yet to delineate
their chemical evolution history.
Therefore, we focus on OHGCs in our following discussions.

Figure~7 shows [Si/Fe], [Ca/Fe], [Ti/Fe], [$\alpha$/Fe], 
and [Fe/H] ratios of the OHGCs as a function of
their {\em current} Galactocentric distance $R_{GC}$.
In the Figure, the [Si/Fe] ratios of the OHGCs appear to
decrease with the Galactocentric distance, while the [Ti/Fe] ratios
increase with the Galactocentric distance,
although the gradient in [Ti/Fe] versus $R_{GC}$ is rather marginal.
[Ca/Fe] and [$\alpha$/Fe] ratios appear to remain constant ($\approx$ +0.30)
with the Galactocentric distance. It is thought that the variations in [Si/Fe]
and [Ti/Fe] ratios with the Galactocentric distance cancel each other,
providing constant [$\alpha$/Fe] ratios with the Galactocentric distance.
The [Fe/H] values of OHGCs show a large scatter around the mean value 
([Fe/H] $\approx$ $-$1.6), however, the absence of gradient in
[Fe/H] versus $R_{GC}$ suggests that sampling different metallicity OHGCs
as a function of $R_{GC}$ is not responsible for the gradients
in [Si/Fe] and [Ti/Fe] ratios with $R_{GC}$.

In Figure~8, we show the [Si/Ti] ratios of the OHGCs as a function of
their Galactocentric distance. In the Figure, we also show 
the least-square fit, the inverse least-square fit, and the bisector liner fit
(Isobe et al.\ 1990) to the data. Our slope of the bisector linear fit to
the data is $\delta$[Si/Ti]/$\delta\log R_{GC}$ = $-$0.765 dex per decade.
We performed a non-parametric Spearman rank-order test (Press et al.\ 1992)
and our results indicate a probability of $\approx$ 0.7~\% that 
the anti-correlation between [Si/Ti] and $R_{GC}$ of the OHGCs is random.
In the Figure, we also show the bisector linear fit to the OHGCs with 
$R_{GC}$ $\leq$ 8 kpc (12 clusters) and we found the slope of 
$\delta$[Si/Ti]/$\delta\log R_{GC}$ = $-$0.976 dex per decade.
We also performed a Spearman rank-order test 
and we obtain a probability of $\approx$ 0.0003~\% that 
the anti-correlation between [Si/Ti] and $R_{GC}$ 
of the OHGCs with $R_{GC}$ $\leq$ 8 kpc is random.\footnote{
It should be noted that the difference in the [Si/Ti] ratios
that we found may be due, in part, to different methods of analyses
and that a thorough re-analysis of all the data using the same
analysis techniques may be very useful.
Also importantly, using [Si~I/Ti~II] ratios to investigate
the [Si/Ti] versus $R_{GC}$ anti-correlation would be very desirable,
since [Si~I/Ti~II] ratios are not sensitive to the uncertainties
in the input model atmosphere as we discussed above.
Unfortunately, most of titanium abundances of the OHGCs
that we adopted from the literature rely on Ti~I lines.}

Figure~7 and 8 suggest that the silicon and titanium
abundances of the OHGCs are related to their current Galactocentric distances.
To explain these elemental abundance gradients,
we show the SNe~II yields as a function of a progenitor mass
with Z = 10$^{-4}$ in Figure~9 (Woosley \& Weaver 1995).
For massive SNe~II models, we adopt the U30B, U35B, and U40B models of
Woosley \& Weaver (1995) following Timmes, Woosley, \& Weaver (1995).
Since Woosley \& Weaver (1995) provided ejected masses of
the individual isotopes from $^1$H to $^{71}$Ge at 2.5$\times$10$^4$ seconds 
after the explosion in their Tables~14A and 14B, 
we need to consider the radioactive decays,
such as negative beta decay and electron capture,
of the short- and intermediate-lived radioactive isotopes to derive
the final products by the SNe~II explosion.
In the inset of the Figure, we also show the SNe~Ia yields by
H\"oeflich, Wheeler, \& Thielemann (1998).
Note that SNe yields are given in units of solar masses
and not in [el/H] and the abscissa of the inset is the atomic number.

As shown in Figure~9, the chemical enrichment by the massive SNe~II
explosion will lead to elemental abundance anomalies, 
in the sense that [O/Fe] and [Si/Fe] ratios will increase,
while [Ti/Fe] ratios\footnote{It should be noted that the SNe~II titanium
yields of Woosley \& Weaver (1995) are too low to explain observed titanium
abundances (Timmes, Woosley, \& Weaver 1995).} will decrease.
We suggest that the gradient in the [Si/Ti] ratios versus $R_{GC}$ 
can be explained by the contributions from the SNe~II explosions with
the different progenitors' masses.
Due to its deeper gravitational potential and the denser environment,
the central regions of our Galaxy would have been better able to retain
the ejecta from massive SNe~II explosions that have higher specific 
kinetic energies compared to the intermediate or the outer halo.
Although weak, the enhanced [O/Fe] ratios of the NGC~6287 RGB stars
with [Na/Fe] $\geq$ 0.0 may support this scenario
(see also Matteucci \& Greggio 1986), since the original [O/Fe] ratios
would have been much higher than the observed ratios
if a substantial amount of oxygen atoms had been already destroyed via
the deep mixing scenario (see, for example, Kraft 1994).
As we have already mentioned, the major contributor to the titanium abundance
due to the SNe~II explosion is $^{48}$Cr, which decays into $^{48}$Ti.
Nakamura et al.\ (1999) argued that the SNe~II Cr yields will decrease 
as the mass of the progenitor increases, leading to decreasing [Cr/Fe] ratios
toward lower [Fe/H]. Thus, the behavior of observed [Cr/Fe] ratios 
in the metal-poor field stars may support the argument that more massive
SNe~II may have contributed to the early nucleosynthesis in our Galaxy.
SNe~Ia explosion models also predict that a significant amount of silicon
can be produced by SNe~Ia explosions
(see Figure~9; see also Nomoto, Thielemann, \& Yokoi 1984), however,
the chemical enrichment by SNe~Ia explosions is not likely responsible for
the gradient in [Si/Ti] ratios versus $R_{GC}$.
The OHGCs are very old to within  0.5 -- 1.0 Gyr
(see, for example, Rosenberg et al.\ 1999;
Lee et al.\ 2001; Lee, Carney, \& Heasley 2002, in preparation)
and SNe~Ia contributions begin to appear at least 1 -- 2 Gyr
after star formation is initiated.

On the other hand, our measurements of $s$-process\footnote{
In the metal-poor field stars, contributions from the $s$-process
can first be seen in metallicities
as low as [Fe/H] $\approx$ $-$2.75 (Burris et al. 2000).}
elements Ba and La may indicate that they formed in PGCCs
which had already been polluted by low-\ or intermediate-mass AGB stars,
unless the observed Ba and La were produced internally and dredged-up
to the stellar surface, and may provide an additional constraint
on the  formation epoch of our program clusters.
If the $s$-process site is intermediate mass AGB stars, which should
begin to appear about 10$^8$ yr after star formation initiated
(Matteucci 2002), the formation timescale of our program clusters
might be as short as a few times 10$^8$ yr after the star formation
is initiated in the Galaxy's central regions.

Perhaps a similar situation can be found in the recent study of
Ivans et al.\ (2001), Fulbright (2001), and Stephens \& Boesgaard (2002).
Ivans et al.\ (2001) compared the intermediate metallicity 
clusters M4 ([Fe/H] = $-$1.08) and M5 ([Fe/H] = $-$1.21) and
they suggested that Si, Ba and La are overabundant in M4 
with respect to those in M5. It should be noted that apogalacticon 
distances of the clusters are $\approx$ 6 kpc for M4 and $\approx$ 40 kpc 
for M5 (Dinescu et al.\ 1999).
Fulbright (2001) performed an abundance study of metal poor field stars
and discussed their kinematics versus elemental abundance relationships.
He suggested that the lower velocity stars 
(mean $R_{max}$ = 9.1 kpc and $\mid Z_{max}\mid$ = 0.6 kpc)
in the solar neighborhood appear to have higher [Si/Fe] and [Ba/Eu] 
ratios than the highest velocity stars 
(mean $R_{max}$ = 37.4 kpc and $\mid Z_{max}\mid$ = 6.2 kpc).
Finally, Stetphens \& Boesgaard (2002) studied elemental abundances
of the kinematically peculiar field halo stars and they argued that
[$\alpha$/Fe] ratios are anticorrelated with apogalacticon distances $R_{apo}$.

On the other hand, the opposite situation can be found in the Galactic 
bulge stars studied by McWilliam \& Rich (1994).
They analyzed the chemical abundances of a dozen metal-rich K giants in
the Galactic bulge and found that [Mg/Fe] and [Ti/Fe] are enhanced
by $\approx$ 0.3 dex, while [Ca/Fe] and [Si/Fe] are depleted by
$\approx$ 0.2 dex. As we discussed above, the oxygen abundances in RGB stars
are hard to determine, due to internal mixing during the RGB phase,
which depletes the oxygen abundance in the stellar photosphere.
The anti-correlation of sodium and oxygen abundances is a good means
by which we may estimate if an individual red giant has experienced
significant mixing or not (Kraft 1994).
In particular, low sodium abundances ([Na/Fe] $\leq$ 0) appears to be
a good indicator of minimal mixing in red giants.
The two stars in McWilliam \& Rich (1994) with [Na/Fe] $\leq$ 0.0 
have a mean [O/Fe] value of $+0.05 \pm 0.05$, consistent
with their result [O/Fe] $\approx$ 0.0.
They also found that the $r$-process element Eu appears to be enhanced,
while the iron-peak elements and $s$-process elements appear to be normal
relative to Fe. Our results and those of McWilliam \& Rich (1994)
may reflect a complexity of the chemical enrichment history near the Galactic
center so that one can hardly draw a boundary between the SNe II contribution
and that due to the SNe Ia events.
Perhaps a comparison between our results and those of McWilliam \& Rich (1994)
might be rather inappropriate since the ages of bulge giants are not known
but presumably they are younger than our program clusters.
During the formation time lag between our program clusters and bulge giants,
the proto-bulge star clouds must have been polluted by the nucleosynthesis
in the Galaxy's central regions and, in particular, 
by infalling gas.\footnote{ Matteucci (2002, see also references therein)
suggested that a fast accumulation of gas in the Galaxy's central regions
accompanied by fast star formation is the best scenario
to reproduce the bulge stars' elemental abundance distribution.}
Therefore, the one-to-one elemental abundance comparison
between our program clusters and bulge giants to trace
the chemical enrichment history of our Galaxy's
central regions may be difficult.

\section{SUMMARY AND CONCLUSIONS}
A chemical abundance study of the old inner halo clusters NGC~6287,
NGC~6293, and NGC~6541 has been presented.
Our metallicity estimates for NGC~6287, NGC~6293, and NGC~6541 are
[Fe/H] = $-$2.01 $\pm$ 0.05, $-$1.99 $\pm$ 0.02, and $-$1.76 $\pm$ 0.02
(internal), respectively, and our metallicity measurements are in good
agreement with previous estimates.

Our radial velocity measurements for NGC~6293 and
NGC~6541 are in good agreement with those by Hesser et al.\ (1986) and
Rutledge et al.\ (1997b), but our radial velocity measurement for NGC~6287
is almost 80 km~s$^{-1}$ larger than that of Hesser et al.\ (1986).

We have discussed that the mean $\alpha$-element abundances 
([$\langle$Si + Ca + Ti$\rangle$/Fe]) of our program clusters
are in good agreement with other globular clusters, 
[$\alpha$/Fe] $\approx$ +0.3,
confirming the previous results of Carney (1996).
However, the individual $\alpha$-elements appear to follow different trends.
The silicon abundances of the inner halo clusters appear to be
enhanced and the titanium abundances appear to be depleted compared
to the intermediate halo clusters. In particular, the [Si/Ti] ratios of
OHGCs appear to be related to their Galactocentric distances,
in the sense that the [Si/Ti] ratios decrease with Galactocentric distance.
We proposed that contributions from different masses of the SNe~II progenitors
that exploded before the formation of OHGCs and
the different initial physical environments surrounding the PGCCs
are responsible for this gradient in [Si/Ti] ratios versus 
Galactocentric distances of OHGCs. 
The high [Si/Ti] ratios toward the Galactic center may be due to
a higher proportion of high-mass SNe~II contributions to the PGCCs.
On the other hand, our program clusters appear have enhanced $s$-process
elemental abundances, providing an additional constraint on the formation
epoch of our program clusters. 
If the $s$-process site is intermediate mass AGB stars, which should
begin to appear about 10$^8$ yr after star formation is initiated,
the formation timescale of our program clusters might be as short
as a few times 10$^8$ yr.

In the future, it would be very desirable to investigate 
the $\alpha$-element abundances and the neutron capture elemental abundances
of the inner halo or bulge clusters with an expanded sample, 
and preferentially at shorter wavelengths
where more neutron capture elements' absorption lines are found.

\acknowledgments
This is part of Ph.D.\ thesis work of J.\ -W.\ Lee at the University
of North Carolina at Chapel Hill.
J.\ -W.\ Lee thanks Chris Sneden, Jim Rose, Art Champagne,
Luisa de Almeida, Mike Habgood, and Inese Ivans for their kind discussions
and Young-Wook Lee and Suk-Jin Yoon for providing recent HB isochrones.
We also thank the anonymous referee for a thorough and beneficial review.
This research was supported by the National Aeronautics and Space
Administration (NASA) grant number GO-07318.04-96A from the Space Telescope
Science Institute, which is operated by the Association of Universities
for Research in Astronomy (AURA), Inc., under NASA contract NAS 5-26555 and
the National Science Foundation grants AST96$-$19381 and AST99$-$88156.
Support for this work was also provided in part by the Creative Research
Initiative Program of Korean Ministry of Science and Technology.

\clearpage

\clearpage

\begin{deluxetable}{clcccccc}
\tablecaption{Journal of observations.}
\tabletypesize{\scriptsize}
\tablenum{1}
\tablewidth{0pc}
\tablehead{
\multicolumn{2}{c}{Id.} &
\multicolumn{1}{c}{$V$} &
\multicolumn{1}{c}{$B-V$} &
\multicolumn{1}{c}{Date/Time} &
\multicolumn{1}{c}{$t_{exp}$} &
\multicolumn{1}{c}{S/N} &
\multicolumn{1}{c}{Note} \\
\colhead{} &
\colhead{} &
\colhead{} &
\colhead{} &
\colhead{(UT Start)} &
\colhead{(sec)} &
\colhead{(total S/N)} }
\startdata
NGC~6287 & 1491 & 14.22 & 1.68 & 1998 Jun 18 23:24 & 3600 & 45 &  \\
         &     &       &      & 1998 Jun 19 00:29 & 3600 & 45 &  \\
         &     &       &      & 1998 Jun 19 01:31 & 3600 & 50 &  \\
         &     & & & 1998 Jun 19 02:32 & 3600 & 50 & S/N $\approx$ 95 \\
         & 1387 & 14.28 & 1.87 & 1998 Jun 20 04:21 & 3600 & 45 &  \\
         &     &       &      & 1998 Jun 22 03:12 & 3000 & 40 &  \\
         &     &       &      & 1998 Jun 22 04:07 & 3000 & 25 &  \\
         &     &       &      & 1999 Jul 05 03:22 & 2400 & 50 &  \\
         &     & & & 1999 Jul 05 04:04 & 2400 & 55 & S/N $\approx$ 95 \\
         & 1191 & 14.58 & 1.64 & 1999 Jul 05 02:26 & 3000 & 40 &  \\
 & & & & & & & \\
NGC~6293 & 2673 & 13.46 & 1.53 & 1998 Jun 21 23:10 & 2400 & 55 &  \\
         &     &       &      & 1998 Jun 21 23:55 & 3600 & 50 &  \\
         &     &       &      & 1998 Jun 22 00:59 & 3600 & 50 &  \\
         &     & & & 1998 Jun 22 02:02 & 3600 & 50 & S/N $\approx$ 100 \\
         & 3857 & 14.11 & 1.49 & 1999 Jul 06 02:46 & 2400 & 35 &  \\
         &     &       &      & 1999 Jul 06 03:29 & 3000 & 35 &  \\
         &     &       &      & 1999 Jul 06 04:21 & 3000 & 45 &  \\
         &     &       &      & 1999 Jul 06 05:14 & 3000 & 45 &  \\
         &     & & & 1999 Jul 06 06:07 & 3000 & 50 &  S/N $\approx$ 95 \\
 & & & & & & & \\
NGC~6541 & I-44 & 12.54 & 1.34 & 1999 Jul 04 01:40 & 2400 & 80 &  \\
         &     &       &      & 1999 Jul 04 02:26 & 2400 & 80 &  \\
         &     & & & 1999 Jul 04 03:07 & 2400 & 80 &  S/N $\approx$ 135 \\
         & II-113 & 12.55 & 1.39 & 1999 Jul 05 04:51 & 2400 & 80 &  \\
         &     &       &      & 1999 Jul 05 05:33 & 2400 & 85 &  \\
         &     & & & 1999 Jul 05 06:14 & 2400 & 80 & S/N $\approx$ 140\\
         & I-21 & 12.46 & 1.31 & 1998 Jun 21 04:45 & 2400 & 70 &  Non Member\\
         &     &       &      & 1998 Jun 21 05:30 & 2400 & 60 &  \\
         &     & & & 1998 Jun 21 06:14 & 2400 & 60 &  S/N $\approx$ 110 \\
\enddata
\end{deluxetable}

\clearpage

\begin{deluxetable}{rcrrcrrrcrrcrr}
\tablecaption{Equivalent widths.}
\tabletypesize{\scriptsize}
\tablenum{2}
\tablewidth{0pc}

\tablehead{
\colhead{$\lambda$ (\AA)} &
\colhead{Elem.} &
\colhead{$\chi$ (eV)} &
\colhead{$\log gf$} &
\colhead{Ref.} &
\multicolumn{3}{c}{NGC~6287} &
\colhead{} &
\multicolumn{2}{c}{NGC~6293} &
\colhead{} &
\multicolumn{2}{c}{NGC~6541} \\

\cline{6-8}\cline{10-11}\cline{13-14}
\colhead{} &
\colhead{} &
\colhead{} &
\colhead{} &
\colhead{} &
\colhead{1491} &
\colhead{1387} &
\colhead{1191} &
\colhead{} &
\colhead{2673} &
\colhead{3857} &
\colhead{} &
\colhead{I-44} &
\colhead{II-113}}

\startdata
6300.23 &  [O I] & 0.00 & $-$9.750 & \OI      &      24 &      27 &      18  && \nodata &       8  &&       7 &       21 \\
6363.88 &  [O I] & 0.02 &$-$10.250 & \OI      &       8 &       9 & \nodata  && \nodata & \nodata  && \nodata &        8 \\

5682.63 &   Na I & 2.10 & $-$0.890 & \NaIa    &      16 &      45 & \nodata  &&      31 & \nodata  && \nodata & \nodata \\
5688.22 &   Na I & 2.10 & $-$0.580 & \NaIa    & \nodata & \nodata & \nodata  &&      51 & \nodata  && \nodata & \nodata \\
6154.23 &   Na I & 2.10 & $-$1.560 & \NaIb    & \nodata &      13 & \nodata  && \nodata & \nodata &&      13 &       10 \\
6160.75 &   Na I & 2.10 & $-$1.260 & \NaIb    & \nodata &      24 & \nodata  &&      16 & 8  &&      23 &       21 \\

5711.10 &   Mg I & 4.34 & $-$1.750 & \MgI     &      33 &      52 & \nodata  &&      16 &      23  && \nodata &      67 \\
7387.69 &   Mg I & 5.75 & $-$1.200 & \MgI     & \nodata &       7 & \nodata  && \nodata & \nodata  && \nodata &      10 \\

6696.03 &   Al I & 3.14 & $-$1.570 & \AlI     & \nodata &      16 & \nodata  &&      19 &      12  &&      54 &      38 \\
6698.67 &   Al I & 3.14 & $-$1.890 & \AlI     & \nodata &       8 & \nodata  &&      11 & \nodata  && \nodata &      21 \\

5665.56 &   Si I & 4.92 & $-$2.040 & \SiIGa   &       8 & \nodata & \nodata  &&      12 & \nodata  && \nodata & \nodata \\
5690.43 &   Si I & 4.93 & $-$1.870 & \SiIGa   &      12 &      14 & \nodata  &&      17 & \nodata  && \nodata & \nodata \\
5701.10 &   Si I & 4.93 & $-$2.050 & \SiIGa   & \nodata &      13 & \nodata  && \nodata & \nodata  && \nodata & \nodata \\
5708.40 &   Si I & 4.95 & $-$1.470 & \SiIGa   & \nodata &      22 & \nodata  &&      27 & \nodata  && \nodata & \nodata \\
5772.15 &   Si I & 5.08 & $-$1.750 & \SiIGa   & \nodata & \nodata & \nodata  &&      16 &      10  &&      21 &      19 \\
5948.55 &   Si I & 5.08 & $-$1.230 & \SiIGa   & \nodata &      30 & \nodata  &&      34 &      21  &&      40 &      44 \\
7034.90 &   Si I & 5.87 & $-$0.880 & \SiIGa   & \nodata &      10 & \nodata  &&      11 & \nodata  &&      17 &      20 \\
7405.77 &   Si I & 5.61 & $-$0.820 & \SiIGa   & \nodata &      20 & \nodata  && \nodata & \nodata  &&      36 &      43 \\

5588.75 &   Ca I & 2.53 &    0.358 & \CaI     &      88 &     105 & \nodata  && \nodata & \nodata  && \nodata & \nodata \\
5590.11 &   Ca I & 2.52 & $-$0.571 & \CaI     &      37 &      48 & \nodata  &&      51 & \nodata  && \nodata & \nodata \\
5594.46 &   Ca I & 2.52 &    0.097 & \CaI     &      83 &      83 & \nodata  &&      99 & \nodata  && \nodata & \nodata \\
5857.45 &   Ca I & 2.93 &    0.240 & \CaI     &      61 &      72 &      58  &&      82 &      57  &&      99 &     103 \\
6161.30 &   Ca I & 2.52 & $-$1.266 & \CaI     &      18 &      18 & \nodata  &&      28 &      13  &&      42 &      45 \\
6166.44 &   Ca I & 2.52 & $-$1.142 & \CaI     &      18 &      22 & \nodata  &&      21 &      21  && \nodata &      47 \\
6169.04 &   Ca I & 2.52 & $-$0.797 & \CaI     &      35 &      46 & \nodata  &&      52 &      34  && \nodata & \nodata \\
6169.56 &   Ca I & 2.53 & $-$0.478 & \CaI     &      48 &      58 & \nodata  &&      62 &      42  &&      83 &      91 \\
6439.08 &   Ca I & 2.53 &    0.390 & \CaI     &     115 &     126 &      96  && \nodata &      98  && \nodata &     146 \\
6449.81 &   Ca I & 2.52 & $-$0.502 & \CaI     &      54 &      62 &      48  &&      68 &      47  &&      84 & \nodata \\
6455.60 &   Ca I & 2.52 & $-$1.290 & \CaI     &      14 &      20 & \nodata  &&      20 &      12  &&      40 &      40 \\
6471.66 &   Ca I & 2.53 & $-$0.686 & \CaI     &      44 &      47 &      37  &&      48 &      43  &&      78 &      88 \\
6493.78 &   Ca I & 2.52 & $-$0.109 & \CaI     &      83 &      91 & \nodata  &&      87 &      64  &&     114 &     118 \\
6499.65 &   Ca I & 2.52 & $-$0.818 & \CaI     &      31 &      48 &      35  &&      39 &      32  &&      71 &      79 \\
6717.68 &   Ca I & 2.71 & $-$0.524 & \CaI     &      48 & \nodata &      29  &&      52 &      37  &&      80 &      91 \\
7148.15 &   Ca I & 2.71 &    0.137 & \CaI     &      89 &     104 &      72  && \nodata &      73  && \nodata &     134 \\
7202.20 &   Ca I & 2.71 & $-$0.262 & \CaI     &      65 &      68 & \nodata  &&      69 &      50  &&     105 &     103 \\

5866.45 &   Ti I & 1.07 & $-$0.784 & \TiIa    &      34 &      42 & \nodata  &&      51 & \nodata  &&      74 &      77 \\
5880.31 &   Ti I & 1.05 & $-$1.989 & \TiIb    & \nodata & \nodata & \nodata  && \nodata & \nodata  &&      13 &      15 \\
5899.30 &   Ti I & 1.05 & $-$1.098 & \TiIa    &      29 &      29 & \nodata  && \nodata & \nodata  && \nodata & \nodata \\
5922.11 &   Ti I & 1.05 & $-$1.410 & \TiIb    & \nodata & \nodata & \nodata  && \nodata & \nodata  && \nodata &      47 \\
5953.16 &   Ti I & 1.89 & $-$0.273 & \TiIc    &      14 &      16 & \nodata  &&      20 &      10  &&      43 &      37 \\
5965.83 &   Ti I & 1.88 & $-$0.353 & \TiIc    &      14 &      12 & \nodata  && \nodata & \nodata  &&      38 &      37 \\
6064.63 &   Ti I & 1.05 & $-$1.888 & \TiIb    & \nodata & \nodata & \nodata  && \nodata & \nodata  &&      20 &      20 \\
6091.18 &   Ti I & 2.30 & $-$0.413 & \TiIc    & \nodata & \nodata & \nodata  && \nodata & \nodata  &&      10 &      14 \\
6126.22 &   Ti I & 1.07 & $-$1.369 & \TiIb    &      17 &      20 & \nodata  && \nodata & \nodata  &&      41 &      44 \\
6258.11 &   Ti I & 1.44 & $-$0.299 & \TiIc    &      37 &      46 & \nodata  &&      39 &      10  &&      71 & \nodata \\
6258.71 &   Ti I & 1.46 & $-$0.270 & \TiIc    &      40 &      43 & \nodata  &&      56 &      23  &&      87 & \nodata \\
6261.11 &   Ti I & 1.43 & $-$0.423 & \TiIc    &      29 &      41 & \nodata  && \nodata &      30  &&      71 & \nodata \\
6312.24 &   Ti I & 1.46 & $-$1.496 & \TiIc    & \nodata & \nodata & \nodata  && \nodata & \nodata  &&      13 &      11 \\
6336.10 &   Ti I & 1.44 & $-$1.687 & \TiIc    & \nodata & \nodata & \nodata  && \nodata & \nodata  &&      10 & \nodata \\
6554.22 &   Ti I & 1.44 & $-$1.162 & \TiIc    & \nodata & \nodata & \nodata  && \nodata & \nodata  &&      25 & \nodata \\
6556.06 &   Ti I & 1.46 & $-$1.018 & \TiIc    & \nodata & \nodata & \nodata  && \nodata & \nodata  && \nodata &      34 \\
6599.13 &   Ti I & 0.90 & $-$2.029 & \TiIb    & \nodata &      10 & \nodata  && \nodata & \nodata  &&      19 &      25 \\
7357.74 &   Ti I & 1.44 & $-$1.066 & \TiIc    & \nodata & \nodata & \nodata  && \nodata & \nodata  &&      36 &      43 \\

6606.95 &  Ti II & 2.06 & $-$2.790 & \TiII    &       8 &       8 & \nodata  && \nodata &       9  &&      16 &      15 \\
7214.74 &  Ti II & 2.59 & $-$1.740 & \TiII    &      16 &      18 & \nodata  &&      18 &      12  &&      33 &      29 \\

5698.52 &    V I & 1.06 & $-$0.111 & \VI      & \nodata & \nodata & \nodata  &&      18 & \nodata  && \nodata & \nodata \\
5703.58 &    V I & 1.05 & $-$0.212 & \VI      & \nodata & \nodata & \nodata  &&      17 &      18  && \nodata & \nodata \\
5727.58 &    V I & 1.08 & $-$0.012 & \VI      & \nodata & \nodata & \nodata  && \nodata & \nodata  &&      47 &      51 \\
5731.25 &    V I & 1.06 & $-$0.730 & \VI      & \nodata & \nodata & \nodata  && \nodata & \nodata  &&      11 &      13 \\
6081.44 &    V I & 1.05 & $-$0.579 & \VI      & \nodata & \nodata & \nodata  &&      10 &       8  &&      15 &      26 \\
6090.22 &    V I & 1.08 & $-$0.062 & \VI      &      17 &      24 & \nodata  &&      23 & \nodata  &&      45 &      44 \\
6111.65 &    V I & 1.04 & $-$0.715 & \VI      & \nodata & \nodata & \nodata  && \nodata & \nodata  &&      15 &      17 \\
6119.53 &    V I & 1.06 & $-$0.320 & \VI      & \nodata & \nodata & \nodata  &&      13 & \nodata  &&      32 &      31 \\
6135.36 &    V I & 1.05 & $-$0.746 & \VI      & \nodata & \nodata & \nodata  && \nodata & \nodata  &&      14 &      17 \\
6251.82 &    V I & 0.29 & $-$1.342 & \VI      &      12 & \nodata & \nodata  &&      15 & \nodata  &&      34 &      36 \\
6274.65 &    V I & 0.27 & $-$1.673 & \VI      & \nodata &      11 & \nodata  && \nodata & \nodata  &&      16 & \nodata \\
6285.15 &    V I & 0.28 & $-$1.510 & \VI      & \nodata & \nodata & \nodata  && \nodata & \nodata  &&      23 &      28 \\
6292.82 &    V I & 0.29 & $-$1.471 & \VI      & \nodata & \nodata & \nodata  &&      14 & \nodata  &&      26 &      33 \\
6296.49 &    V I & 0.30 & $-$1.590 & \VI      & \nodata & \nodata & \nodata  && \nodata & \nodata  &&      27 &      31 \\

5783.89 &   Cr I & 3.32 & $-$0.295 & \CrI     & \nodata & \nodata & \nodata  && \nodata & \nodata  &&      18 & \nodata \\
6978.49 &   Cr I & 3.46 &    0.142 & \CrI     & \nodata & \nodata & \nodata  && \nodata & \nodata  && \nodata &      22 \\
7355.93 &   Cr I & 2.89 & $-$0.285 & \CrI     &      12 &      18 & \nodata  &&      16 &      11  && \nodata &      38 \\
7400.23 &   Cr I & 2.90 & $-$0.111 & \CrI     &      18 &      28 & \nodata  &&      24 &      15  &&      66 &      51 \\

5569.62 &   Fe I & 3.42 & $-$0.544 & \FeIBKb  &      82 & \nodata & \nodata  &&     100 & \nodata  && \nodata & \nodata \\
5586.76 &   Fe I & 3.37 & $-$0.161 & \FeIOB   & \nodata &     123 & \nodata  &&     117 & \nodata  && \nodata & \nodata \\
5701.55 &   Fe I & 2.56 & $-$2.216 & \FeId    &      53 &      76 & \nodata  &&      65 & \nodata  && \nodata & \nodata \\
5705.47 &   Fe I & 4.30 & $-$1.421 & \FeIBKb  & \nodata & \nodata & \nodata  &&       9 & \nodata  &&      11 &      12 \\
5753.12 &   Fe I & 4.26 & $-$0.705 & \FeIOB   &      19 &      26 & \nodata  &&      30 &      19  &&      43 &      48 \\
5775.08 &   Fe I & 4.22 & $-$1.314 & \FeIOB   &      11 &      11 & \nodata  &&      12 &      10  &&      27 &      26 \\
5778.45 &   Fe I & 2.59 & $-$3.475 & \FeIBKb  & \nodata &      14 & \nodata  &&      11 & \nodata  &&      16 &      21 \\
5909.97 &   Fe I & 3.21 & $-$2.643 & \FeIBKb  &       9 & \nodata & \nodata  && \nodata & \nodata  &&      26 & \nodata \\
5916.25 &   Fe I & 2.45 & $-$2.994 & \FeIc    &      24 &      48 &      22  &&      39 &      20  &&      65 &      62 \\
5956.69 &   Fe I & 0.86 & $-$4.608 & \FeIe    &      65 &      84 &      36  &&      81 &      47  &&      93 &     106 \\
6027.05 &   Fe I & 4.08 & $-$1.106 & \FeIOB   &      15 &      27 &      16  &&      19 &      18  &&      37 &      41 \\
6065.48 &   Fe I & 2.61 & $-$1.530 & \FeId    &     105 &     112 &      86  &&     114 &      94  &&     132 &     136 \\
6082.71 &   Fe I & 2.22 & $-$3.573 & \FeIc    &      13 &      21 &      11  &&      25 &      12  &&      43 &      42 \\
6137.69 &   Fe I & 2.59 & $-$1.403 & \FeId    &     108 &     134 & \nodata  &&     130 &      97  && \nodata & \nodata \\
6151.62 &   Fe I & 2.18 & $-$3.299 & \FeIc    &      29 &      47 &      30  &&      48 &      32  &&      65 &      68 \\
6165.36 &   Fe I & 4.14 & $-$1.490 & \FeIOB   &      11 &      12 & \nodata  &&      11 & \nodata  &&      18 &      15 \\
6173.34 &   Fe I & 2.22 & $-$2.880 & \FeIc    &      52 &      65 &      38  &&      61 &      42  &&      93 &      90 \\
6200.31 &   Fe I & 2.61 & $-$2.437 & \FeId    &      54 &      64 &      33  &&      57 &      38  &&      81 &      84 \\
6219.28 &   Fe I & 2.20 & $-$2.433 & \FeIc    &      87 &     101 &      65  &&     104 &      75  &&     114 &     106 \\
6229.23 &   Fe I & 2.85 & $-$2.846 & \FeIBKa  &      12 &      17 &      11  &&      16 &      14  &&      35 &      35 \\
6230.73 &   Fe I & 2.56 & $-$1.281 & \FeId    &     116 &     150 & \nodata  &&     155 &     109  && \nodata & \nodata \\
6232.64 &   Fe I & 3.65 & $-$1.283 & \FeIBKb  &      34 &      41 &      28  &&      38 &      27  &&      57 &      61 \\
6240.64 &   Fe I & 2.22 & $-$3.190 & \FeIBKa  &      35 &      41 & \nodata  &&      39 &      20  &&      65 &      63 \\
6246.32 &   Fe I & 3.60 & $-$0.894 & \FeIBKa  &      65 &      66 &      41  &&      67 &      56  &&      94 &      95 \\
6252.55 &   Fe I & 2.40 & $-$1.687 & \FeIc    &     111 &     119 &      93  &&     129 &      92  &&     138 &     143 \\
6265.13 &   Fe I & 2.18 & $-$2.550 & \FeIc    &      72 &      96 &      73  &&      97 &      73  &&     112 &     116 \\
6270.22 &   Fe I & 2.86 & $-$2.505 & \FeIBKa  &      34 &      29 & \nodata  &&      28 &      22  &&      53 &      52 \\
6280.63 &   Fe I & 0.86 & $-$4.390 & \FeIe    &      76 &     108 & \nodata  &&     106 &      70  && \nodata & \nodata \\
6297.79 &   Fe I & 2.22 & $-$2.740 & \FeIc    &      56 &      81 &      50  &&      78 &      42  &&      96 &     104 \\
6301.50 &   Fe I & 3.65 & $-$0.766 & \FeIBKa  &      65 & \nodata &      54  &&      71 &      59  &&      89 &      97 \\
6322.69 &   Fe I & 2.59 & $-$2.426 & \FeId    &      50 &      67 &      38  &&      66 &      36  && \nodata &      85 \\
6335.33 &   Fe I & 2.20 & $-$2.194 & \FeIOB   & \nodata &     110 &      80  &&     114 &      85  &&     129 &     130 \\
6336.82 &   Fe I & 3.69 & $-$0.916 & \FeIBKb  &      54 &      62 &      37  &&      59 &      40  &&      87 &      85 \\
6344.15 &   Fe I & 2.43 & $-$2.923 & \FeIc    &      41 &      50 &      38  &&      56 &      27  &&      79 &      71 \\
6358.69 &   Fe I & 0.86 & $-$4.468 & \FeIb    &      78 &     109 & \nodata  &&     101 &      60  && \nodata & \nodata \\
6393.60 &   Fe I & 2.43 & $-$1.469 & \FeIBKa  &     120 &     135 & \nodata  &&     144 &     119  &&     153 &     151 \\
6408.02 &   Fe I & 3.69 & $-$1.066 & \FeIBKa  &      41 &      51 &      35  &&      50 &      34  &&      72 &      77 \\
6411.65 &   Fe I & 3.65 & $-$0.734 & \FeIBKa  &      65 &      82 &      57  &&      77 &      53  && \nodata &     100 \\
6421.35 &   Fe I & 2.28 & $-$2.027 & \FeIc    &     101 &     120 & \nodata  &&     123 &      85  && \nodata & \nodata \\
6430.84 &   Fe I & 2.18 & $-$2.006 & \FeIc    &     113 &     134 &     104  &&     137 &     102  &&     138 &     148 \\
6481.87 &   Fe I & 2.28 & $-$2.984 & \FeIc    &      42 &      62 &      46  &&      56 &      33  &&      78 &      80 \\
6494.98 &   Fe I & 2.40 & $-$1.273 & \FeIc    &     137 &     158 &     121  &&     162 &     129  &&     173 &     188 \\
6498.95 &   Fe I & 0.96 & $-$4.687 & \FeIe    &      56 &      71 & \nodata  &&      77 &      39  && \nodata & \nodata \\
6574.23 &   Fe I & 0.99 & $-$5.004 & \FeIe    &      35 &      41 &      22  &&      52 &      26  &&      59 &      74 \\
6581.21 &   Fe I & 1.48 & $-$4.708 & \FeIBKa  &      16 &      27 & \nodata  &&      24 & \nodata  &&      35 &      38 \\
6592.91 &   Fe I & 2.73 & $-$1.490 & \FeIOB   &      97 &     112 &      94  &&     112 &      80  &&     129 &     134 \\
6593.87 &   Fe I & 2.43 & $-$2.422 & \FeIc    &      72 &      84 &      56  &&      84 &      58  &&     100 &     102 \\
6609.11 &   Fe I & 2.56 & $-$2.692 & \FeId    &      37 &      54 &      23  &&      49 &      28  &&      78 &      81 \\
6625.02 &   Fe I & 1.01 & $-$5.366 & \FeIe    &      13 &      26 & \nodata  &&      23 &      11  &&      42 &      45 \\
6663.44 &   Fe I & 2.42 & $-$2.479 & \FeIc    &      69 &      85 & \nodata  &&      84 & \nodata  && \nodata & \nodata \\
6677.99 &   Fe I & 2.69 & $-$1.435 & \FeIOB   &     110 &     129 &      94  &&     120 &      95  &&     140 &     144 \\
6750.15 &   Fe I & 2.42 & $-$2.621 & \FeIc    &      56 &      72 &      44  &&      74 &      50  &&      97 &      89 \\
6752.70 &   Fe I & 4.64 & $-$1.273 & \FeIBKb  & \nodata & \nodata & \nodata  && \nodata & \nodata  &&       8 &       8 \\
6810.26 &   Fe I & 4.61 & $-$1.003 & \FeIOB   & \nodata &       9 & \nodata  && \nodata & \nodata  &&      15 &      18 \\
6945.20 &   Fe I & 2.42 & $-$2.482 & \FeIc    &      73 &      87 &      57  &&      91 &      54  &&     113 &     119 \\
7112.17 &   Fe I & 2.99 & $-$3.044 & \FeIBKb  &      10 &      12 & \nodata  &&      12 & \nodata  &&      24 &      27 \\
7189.15 &   Fe I & 3.07 & $-$2.825 & \FeIBKb  &       9 & \nodata & \nodata  && \nodata & \nodata  &&      28 & \nodata \\
7401.69 &   Fe I & 4.19 & $-$1.644 & \FeIBKb  & \nodata & \nodata & \nodata  && \nodata & \nodata  &&      16 &      20 \\
7511.02 &   Fe I & 4.18 &    0.082 & \FeIOB   &      79 &      92 &      66  &&      89 &      71  &&     110 &     114 \\
7723.20 &   Fe I & 2.95 & $-$3.617 & \FeIc    &      18 &      34 & \nodata  && \nodata & \nodata  && \nodata & \nodata \\

5991.38 &  Fe II & 3.15 & $-$3.557 & \FeII    & \nodata & \nodata & \nodata  &&      14 &      11  &&      17 &      20 \\
6084.11 &  Fe II & 3.20 & $-$3.808 & \FeII    & \nodata & \nodata & \nodata  && \nodata & \nodata  &&      10 &      11 \\
6149.26 &  Fe II & 3.89 & $-$2.724 & \FeII    &      12 &      14 &      12  &&      13 &      13  &&      19 &      14 \\
6247.56 &  Fe II & 3.89 & $-$2.329 & \FeII    &      22 &      22 &      20  &&      20 &      24  &&      34 &      33 \\
6416.92 &  Fe II & 3.89 & $-$2.740 & \FeII    &      13 &      13 &      11  &&      11 & \nodata  && \nodata &      20 \\
6432.68 &  Fe II & 2.89 & $-$3.708 & \FeII    &      21 &      24 &      19  &&      20 &      19  &&      32 &      32 \\
6456.38 &  Fe II & 3.90 & $-$2.075 & \FeII    &      32 &      37 &      31  &&      39 &      36  &&      45 &      47 \\

6093.14 &   Co I & 1.74 & $-$2.440 & \CoI     & \nodata & \nodata & \nodata  && \nodata & \nodata  &&      17 &      16 \\
6117.00 &   Co I & 1.79 & $-$2.490 & \CoI     & \nodata & \nodata & \nodata  && \nodata & \nodata  &&      12 &      12 \\
6189.00 &   Co I & 1.71 & $-$2.450 & \CoI     & \nodata & \nodata & \nodata  && \nodata & \nodata  &&      16 &      17 \\
6632.43 &   Co I & 2.28 & $-$2.000 & \CoI     & \nodata & \nodata & \nodata  && \nodata & \nodata  &&       9 & \nodata \\
6771.04 &   Co I & 1.88 & $-$1.970 & \CoI     & \nodata & \nodata & \nodata  && \nodata & \nodata  &&      34 &      32 \\
6814.94 &   Co I & 1.96 & $-$1.900 & \CoI     & \nodata &      11 & \nodata  &&      11 & \nodata  &&      29 &      26 \\
7052.87 &   Co I & 1.96 & $-$1.620 & \CoI     &      23 &      24 & \nodata  &&      24 &      13  &&      52 &      50 \\
7054.05 &   Co I & 2.72 & $-$1.530 & \CoI     & \nodata & \nodata & \nodata  && \nodata & \nodata  &&       9 & \nodata \\
7417.37 &   Co I & 2.04 & $-$2.070 & \CoI     & \nodata &       9 & \nodata  && \nodata & \nodata  && \nodata & \nodata \\

5748.35 &   Ni I & 1.68 & $-$3.260 & \NiI     & \nodata & \nodata & \nodata  &&      21 & \nodata  &&      33 &      38 \\
5754.66 &   Ni I & 1.94 & $-$2.330 & \NiI     & \nodata &      56 & \nodata  &&      49 &      37  &&      82 & \nodata \\
5892.87 &   Ni I & 4.15 & $-$2.350 & \NiI     & \nodata &      67 & \nodata  && \nodata & \nodata  && \nodata & \nodata \\
6108.11 &   Ni I & 1.68 & $-$2.450 & \NiI     &      54 & \nodata & \nodata  &&      64 &      36  && \nodata &      84 \\
6175.36 &   Ni I & 4.09 & $-$0.530 & \NiI     & \nodata & \nodata & \nodata  && \nodata & \nodata  &&      18 &      15 \\
6176.81 &   Ni I & 4.08 & $-$0.530 & \NiI     & \nodata & \nodata & \nodata  && \nodata & \nodata  &&      19 &      23 \\
6177.24 &   Ni I & 4.24 & $-$3.500 & \NiI     & \nodata &      11 & \nodata  && \nodata & \nodata  &&      18 &      14 \\
6256.35 &   Ni I & 1.68 & $-$2.480 & \NiI     &      69 &      84 &      53  && \nodata & \nodata  &&      92 &     107 \\
6327.59 &   Ni I & 1.68 & $-$3.150 & \NiI     &      28 &      29 & \nodata  &&      34 & \nodata  &&      51 &      52 \\
6378.25 &   Ni I & 4.15 & $-$0.890 & \NiI     & \nodata & \nodata & \nodata  && \nodata & \nodata  && \nodata &       9 \\
6482.80 &   Ni I & 1.94 & $-$2.630 & \NiI     & \nodata & \nodata & \nodata  &&      31 & \nodata  && \nodata &      49 \\
6586.31 &   Ni I & 1.95 & $-$2.810 & \NiI     &      24 &      25 & \nodata  &&      24 &      15  &&      49 &      47 \\
6643.63 &   Ni I & 1.68 & $-$2.300 & \NiI     &      79 &      99 &      60  &&      97 &      58  &&     114 &     122 \\
6767.77 &   Ni I & 1.83 & $-$2.170 & \NiI     &      73 & \nodata &      50  &&      85 &      59  &&      96 &     101 \\
6914.56 &   Ni I & 1.95 & $-$2.270 & \NiI     &      53 &      68 & \nodata  &&      62 & \nodata  &&      89 &      92 \\
7122.19 &   Ni I & 3.54 &    0.040 & \NiI     & \nodata &      68 &      49  && \nodata & \nodata  &&      89 &      91 \\
7197.01 &   Ni I & 1.94 & $-$2.680 & \NiI     &      38 & \nodata & \nodata  && \nodata & \nodata  &&      66 &      86 \\
7261.92 &   Ni I & 1.95 & $-$2.700 & \NiI     & \nodata & \nodata & \nodata  && \nodata &      31  && \nodata & \nodata \\
7414.50 &   Ni I & 1.99 & $-$2.570 & \NiI     &      37 &      54 &      25  && \nodata &      27  &&      71 &      73 \\
7422.28 &   Ni I & 3.63 & $-$0.140 & \NiI     &      43 &      50 &      31  && \nodata & \nodata  &&      74 &      73 \\
7714.31 &   Ni I & 1.94 & $-$2.200 & \NiI     &      72 &      89 &      46  && \nodata & \nodata  && \nodata & \nodata \\

5853.69 &  Ba II & 0.60 & $-$1.010 & \BaII    &      92 &     114 &      97  &&      95 &      75  &&     111 &     118 \\
6141.73 &  Ba II & 0.70 & $-$0.080 & \BaII    &     147 &     164 &     136  &&     141 &     117  &&     158 &     169 \\
6496.91 &  Ba II & 0.60 & $-$0.380 & \BaII    &     154 &     181 &     142  &&     150 &     114  &&     162 &     178 \\

6390.48 &  La II & 0.32 & $-$1.450 & \LaII    &      11 &      16 & \nodata  &&      10 & \nodata  &&      17 &      17 \\
6774.27 &  La II & 0.13 & $-$1.820 & \LaII    &      11 &      13 & \nodata  &&      11 & \nodata  &&      14 &      18 \\

6645.06 &  Eu II & 1.38 &    0.204 & \EuII    &      11 &      16 & \nodata  &&      12 &      10  &&      16 &      17 \\
7217.55 &  Eu II & 1.23 & $-$0.301 & \EuII    &       7 &       8 & \nodata  &&       9 & \nodata  &&       9 & \nodata \\

\enddata
\end{deluxetable}

\clearpage

\begin{deluxetable}{cll}
\tabletypesize{\scriptsize}
\tablecaption{References for $\log gf$ values}
\tablenum{3}
\tablewidth{0pc}

\tablehead{
\colhead{Ref. No. } & \colhead{Reference} &
\colhead{Elem.}}
\startdata
\OI       & Lambert (1978)   & [O I] \\
\NaIa     & Prochaska et al. (2000) & Na I \\
\NaIb     & Kraft et al. (1992)     & Na I \\
\MgI      & Th\'evenin (1990)  & Mg I \\
\AlI      & Sneden et al. (1997)    & Al I \\
\SiIGa    & Garz (1973)      & Si I \\
\CaI      & Smith (1981)     & Ca I \\
\TiIa     & Blackwell et al. (1982a)    & Ti I \\
\TiIb     & Blackwell et al. (1983)     & Ti I \\
\TiIc     & Blackwell et al. (1986b)     & Ti I \\
\TiII     & Fuhr et al. (1988a)     & Ti II, Cr I \\
\VI       & Whaling et al. (1985)   & V I \\
\FeIb     & Blackwell et al. (1979)    & Fe I \\
\FeIc     & Blackwell et al. (1982b)    & Fe I \\
\FeId     & Blackwell et al. (1982c)    & Fe I \\
\FeIe     & Blackwell et al. (1986a)     & Fe I \\
\FeIOB    & O'Brian et al. (1991)    & Fe I \\
\FeIBKa   & Bard et al. (1991)        & Fe I \\
\FeIBKb   & Bard \& Kock (1994)        & Fe I \\
\FeII     & Bi\'emont (1991)   & Fe II \\
\CoI      & Cardon (1982)    & Co I \\
\NiI      & Fuhr et al. (1988b)     & Ni I \\
\BaII     & Sneden et al. (1997) & Ba II \\
\LaII     & Kurucz (1995) & La II \\
\EuII     & Bi\'emont et al. (1982) & Eu II\\
\enddata
\end{deluxetable}

\clearpage

\begin{deluxetable}{lccccccccccc}
\tabletypesize{\scriptsize}
\tablecaption{Model atmosphere parameters.}
\tablenum{4}
\tablewidth{6.5in}
\scriptsize
\tablehead{
\multicolumn{3}{c}{} &
\multicolumn{2}{c}{Photometric} &
\multicolumn{1}{c}{} &
\multicolumn{3}{c}{Spectroscopic} &
\multicolumn{1}{c}{} &
\multicolumn{2}{c}{$\Delta$(Ph. $-$ Sp.)} \\
\cline{4-5}\cline{7-9}\cline{11-12}
\colhead{~~~~~Id} &
\colhead{$E(B-V)$} &
\colhead{$(m-M)_0$} &
\colhead{$T_{eff}$} &
\colhead{$\log g$} & &
\colhead{$T_{eff}$} &
\colhead{$\log g$} &
\colhead{$v_{turb}$} & &
\colhead{$\Delta T_{eff}$} &
\colhead{$\Delta\log g$} \\
\multicolumn{3}{c}{} &
\multicolumn{1}{c}{(K)} &
\multicolumn{1}{c}{(0.8$M_\odot$)} &
\multicolumn{1}{c}{} &
\multicolumn{1}{c}{(K)} &
\multicolumn{1}{c}{} &
\multicolumn{1}{c}{(km/s)} &
\multicolumn{1}{c}{} &
\multicolumn{1}{c}{(K)} &
\multicolumn{1}{c}{} }
\startdata
NGC~6287-1491  &0.62&14.35&4443&1.0 &&4375&    0.7&1.75&& ~~68&0.3\\
NGC~6287-1387  &0.62&14.35&4219&0.8 &&4250&    0.5&1.90&& $-$31&0.3\\
NGC~6287-1191  &0.62&14.35&4493&1.1 &&\nodata& 0.9&1.80&&\nodata&0.2\\
NGC~6293-2673  &0.40&14.61&4356&0.8 &&4250&    0.5&1.90&& ~106&0.3\\
NGC~6293-3857  &0.40&14.61&4405&1.1 &&4450&    0.7&1.75&&$-$45&0.4 \\
NGC~6541-I-44  &0.14&14.19&4271&0.8 &&4250&    0.7&1.85&&~~21&0.1\\
NGC~6541-II-113&0.14&14.19&4212&0.8 &&4200&    0.5&1.80&&~~11&0.1\\
\enddata
\end{deluxetable}

\clearpage

\begin{deluxetable}{clccc}
\tabletypesize{\scriptsize}
\tablecaption{Radial velocity measurements}
\tablenum{5}
\tablewidth{0pc}

\tablehead{
\multicolumn{2}{c}{Id.} &
\multicolumn{1}{c}{$v_r$} &
\multicolumn{1}{c}{$\sigma$} &
\multicolumn{1}{c}{Note} \\
\colhead{} & \colhead{} &
\colhead{(km s$^{-1}$)} &
\colhead{(km s$^{-1}$)} &
\colhead{}}
\startdata
NGC~6287 & 1491 & $-$290.04 & 0.94 & \\
         & 1387 & $-$294.06 & 0.89 & \\
         & 1191 & $-$282.31 & 0.67 & \\
NGC~6293 & 2673 & $-$158.57 & 0.89 & \\
         & 3857 & $-$145.27 & 0.69 & \\
NGC~6541 & I-44 & $-$162.52 & 0.43 & \\
         & II-113 & $-$172.49 & 0.93 & \\
         & I-21 & ~$-$34.13 & 0.79 & Non Member \\
\enddata
\end{deluxetable}

\clearpage

\begin{deluxetable}{lcccrccccrccccr}
\tabletypesize{\scriptsize}
\tablecaption{Elemental abundances of NGC~6287.}
\tablenum{6}
\tablewidth{0pc}
\tablehead{
\multicolumn{1}{c}{} & \multicolumn{4}{c}{1491} & \multicolumn{1}{c}{} &
\multicolumn{4}{c}{1387} & \multicolumn{1}{c}{} & \multicolumn{4}{c}{1191} \\
\cline{2-5}\cline{7-10}\cline{12-15}
\colhead{}  & \colhead{$\log g_{P}$} & \colhead{$\log g_{S}$} &
\colhead{$\sigma$} & \colhead{n} & \colhead{} & \colhead{$\log g_{P}$} &
\colhead{$\log g_{S}$} & \colhead{$\sigma$} & \colhead{n} & \colhead{} &
\colhead{$\log g_{P}$} & \colhead{$\log g_{S}$} & \colhead{$\sigma$} &
\colhead{n}}
\startdata
[Fe/H]$_{\mathrm{I}}$&$-$2.17&$-$2.15&0.09& 53 &&$-$2.14&$-$2.10 &0.09& 53 &&$-$2.18&$-$2.16&0.12& 33 \\

[Fe/H]$_{\mathrm{II}}$&$-$1.99&$-$2.11&0.05& 5 &&$-$1.95&$-$2.09&0.06& 5 &&$-$2.08&$-$2.16&0.07 & 5 \\

[Fe/H]$_{avg}$&\nodata&$-$2.13&0.10&\nodata &&\nodata&$-$2.10&0.11&  &&\nodata& $-$2.16&0.04&\nodata  \\

& & & & & & & & & & & & &\\

[O/Fe]         &~~0.34&  ~~0.37 &    0.01 & 2 &&~~0.24&  ~~0.28 & 0.01 & 2 &&~~0.34&  ~~0.39 & \nodata & 1 \\

[Na/Fe]        &~~0.28&  ~~0.26 & \nodata & 1 &&~~0.71&  ~~0.72 & 0.03 & 3 &&\nodata& \nodata & \nodata & \nodata\\

[Mg/Fe]        &~~0.26&  ~~0.24 & \nodata & 1 &&~~0.44&  ~~0.44 & 0.01 & 2 && \nodata & \nodata & \nodata & \nodata \\

[Al/Fe]        &\nodata & \nodata & \nodata & \nodata &&~~0.93&  ~~0.88 & 0.00 & 2 &&\nodata & \nodata & \nodata & \nodata \\

[Si/Fe]        &~~0.57&  ~~0.51 &    0.05 & 2 &&~~0.52&  ~~0.46 & 0.05 & 6 &&\nodata & \nodata & \nodata & \nodata \\

[Ca/Fe]        &~~0.22&  ~~0.21 &    0.10 & 17 &&~~0.16&  ~~0.17 & 0.08 & 16 &&~~0.24&  ~~0.17 &    0.06 & 7 \\

[Ti/Fe]$_{\mathrm{I}}$  &~~0.19&  ~~0.18 & 0.08 & 8 &&~~0.02&  ~~0.06 & 0.06 & 9 &&\nodata & \nodata & \nodata & \nodata \\

[Ti/Fe]$_{\mathrm{II}}$ &~~0.16&  ~~0.18 & 0.05 & 2 &&~~0.11&  ~~0.10 & 0.01 & 2 &&\nodata & \nodata & \nodata & \nodata \\

[V/Fe]         &$-$0.02& $-$0.03 &    0.01 & 3 &&$-$0.09& $-$0.05 & 0.04 & 2 && \nodata &\nodata & \nodata & \nodata \\

[Cr/Fe]        &$-$0.14& $-$0.16 &    0.02 & 2 &&$-$0.11& $-$0.10 & 0.04 & 2 && \nodata &\nodata & \nodata & \nodata \\

[Co/Fe]        &~~0.27&  ~~0.25 & \nodata & 1 &&~~0.07&  ~~0.06 & 0.07 & 3 && \nodata &\nodata & \nodata & \nodata \\

[Ni/Fe]        &~~0.09&  ~~0.05 &    0.09 & 11 &&~~0.04&  ~~0.01 & 0.12 & 12 &&$-$0.02& $-$0.03 &    0.10 & 7 \\

[Ba/Fe]$_{\mathrm{II}}$ &~~0.34&  ~~0.38 &    0.18 & 3 &&~~0.43&  ~~0.49 & 0.16 & 3 &&~~0.39&  ~~0.40 &    0.09 & 3 \\

[La/Fe]$_{\mathrm{II}}$ &~~0.30&  ~~0.32 &    0.06 & 2 &&~~0.26&  ~~0.30 & 0.02 & 2 && \nodata &\nodata & \nodata & \nodata \\

[Eu/Fe]$_{\mathrm{II}}$ &~~0.45&  ~~0.47 &    0.02 & 2 &&~~0.44&  ~~0.47 & 0.04 & 2 &&\nodata& \nodata & \nodata & \nodata \\
\enddata
\end{deluxetable}

\clearpage

\begin{deluxetable}{lcccrccccr}
\tabletypesize{\scriptsize}
\tablecaption{Elemental abundances of NGC~6293.}
\tablenum{7}
\tablewidth{0pc}
\tablehead{
\multicolumn{1}{c}{} & \multicolumn{4}{c}{2673} &
\multicolumn{1}{c}{} & \multicolumn{4}{c}{3857} \\
\cline{2-5}\cline{7-10}
\colhead{}  & \colhead{$\log g_{P}$} & \colhead{$\log g_{S}$} &
\colhead{$\sigma$} & \colhead{n} & \colhead{} & \colhead{$\log g_{P}$} &
\colhead{$\log g_{S}$} & \colhead{$\sigma$} & \colhead{n}}
\startdata
[Fe/H]$_{\mathrm{I}}$   &$-$2.16&  $-$2.16 & 0.10 & 54 &&$-$2.20& $-$2.18 & 0.09 & 45 \\

[Fe/H]$_{\mathrm{II}}$  &$-$1.99&  $-$2.12 & 0.08 & 6 && $-$1.99&$-$2.15 & 0.04 & 5 \\

[Fe/H]$_{avg}$ & \nodata & $-$2.10 & 0.14 &\nodata &&\nodata& $-$2.17 & 0.10 &\nodata \\

& & & & & & & & & \\

[O/Fe]         & \nodata& \nodata & \nodata & \nodata && $-$0.14& $-$0.09 & \nodata & 1 \\

[Na/Fe]        &~~0.40&   ~~0.41 &    0.06 & 2 &&~~0.38&  ~~0.40 & \nodata & 1 \\

[Mg/Fe]        &$-$0.26&  $-$0.25 & \nodata & 1 &&~~0.17&  ~~0.14 & \nodata & 1 \\

[Al/Fe]        &~~1.01&   ~~1.01 &    0.04 & 2 &&~~0.97&  ~~0.97 & \nodata & 1 \\

[Si/Fe]        &~~0.68&   ~~0.61 &    0.06 & 6 &&~~0.54&  ~~0.51 &    0.10 & 2 \\

[Ca/Fe]        &~~0.22&   ~~0.23 &    0.10 & 14 &&~~0.23&  ~~0.24 &    0.07 & 14 \\

[Ti/Fe]$_{\mathrm{I}}$  &~~0.08&   ~~0.09 &    0.11 & 4 &&~~0.12&  ~~0.14 &    0.08 & 4 \\

[Ti/Fe]$_{\mathrm{II}}$ &~~0.09&   ~~0.13 & \nodata & 1 &&~~0.14&  ~~0.16 &    0.21 & 2 \\

[V/Fe]         &$-$0.15&  $-$0.15 &    0.06 & 7 &&~~0.28&  ~~0.30 &    0.03 & 2 \\

[Cr/Fe]        &$-$0.17&  $-$0.17 &    0.03 & 2 &&$-$0.10& $-$0.09 &    0.01 & 2 \\

[Co/Fe]        &~~0.06&   ~~0.03 &    0.05 & 2 &&~~0.15&  ~~0.15 & \nodata & 1 \\

[Ni/Fe]        &$-$0.04&  $-$0.07 &    0.09 & 9 &&$-$0.03& $-$0.03 &    0.09 & 7 \\

[Ba/Fe]$_{\mathrm{II}}$ &~~0.07&   ~~0.16 &    0.09 & 3 &&$-$0.07& $-$0.07 &    0.01 & 3 \\

[La/Fe]$_{\mathrm{II}}$ &~~0.15&   ~~0.18 &    0.08 & 2 && \nodata & \nodata & \nodata & \nodata \\

[Eu/Fe]$_{\mathrm{II}}$ &~~0.43&   ~~0.46 &    0.08 & 2 &&~~0.41&  ~~0.45 & \nodata & 1 \\
\enddata
\end{deluxetable}

\clearpage

\begin{deluxetable}{lcccrccccr}
\tabletypesize{\scriptsize}
\tablecaption{Elemental abundances of NGC~6541.}
\tablenum{8}
\tablewidth{0pc}
\tablehead{
\multicolumn{1}{c}{} & \multicolumn{4}{c}{I-44} &
\multicolumn{1}{c}{} & \multicolumn{4}{c}{II-113} \\
\cline{2-5}\cline{7-10}
\colhead{}  & \colhead{$\log g_{P}$} & \colhead{$\log g_{S}$} &
\colhead{$\sigma$} & \colhead{n} & \colhead{} & \colhead{$\log g_{P}$} &
\colhead{$\log g_{S}$} & \colhead{$\sigma$} & \colhead{n}}
\startdata
[Fe/H]$_{\mathrm{I}}$   &$-$1.85&  $-$1.85 & 0.08 & 47 &&$-$1.86& $-$1.86 &  0.08 & 47  \\

[Fe/H]$_{\mathrm{II}}$  &$-$1.79&  $-$1.83 & 0.07 & 6 &&$-$1.72& $-$1.85 &  0.08 & 7 \\

[Fe/H]$_{avg}$ & \nodata& $-$1.84 & 0.11 & &&\nodata& $-$1.86 &  0.11  & \\

 & & & & & & & & &\\

[O/Fe]         &$-$0.58&  $-$0.55 & \nodata & 1 && $-$0.09&$-$0.07 &    0.05 & 2 \\

[Na/Fe]        &~~0.40&   ~~0.40 &    0.00 & 2 &&~~0.26&  ~~0.29 &    0.04 & 2 \\

[Mg/Fe]        &\nodata &  \nodata & \nodata & \nodata &&~~0.34&  ~~0.36 &    0.02 & 2 \\

[Al/Fe]        &~~1.33&   ~~1.30 & \nodata & 1 &&~~1.02&  ~~1.05 &    0.01 & 2 \\

[Si/Fe]        &~~0.51&   ~~0.51 &    0.06 & 4 &&~~0.52&  ~~0.56 &    0.08 &  4 \\

[Ca/Fe]        &~~0.31&   ~~0.32 &    0.08 & 10 &&~~0.36&  ~~0.39 &    0.09 & 11 \\

[Ti/Fe]$_{\mathrm{I}}$  &~~0.22&   ~~0.22 &  0.07 & 15 &&~~0.17&  ~~0.20 &    0.06 & 12 \\

[Ti/Fe]$_{\mathrm{II}}$ &~~0.27&   ~~0.28 &  0.01 & 2 &&~~0.17&  ~~0.19 &    0.02 & 2 \\

[V/Fe]         &$-$0.09&  $-$0.09 &    0.07 & 12 &&$-$0.10& $-$0.07 &    0.07 & 11 \\

[Cr/Fe]        &~~0.25&   ~~0.25 &    0.01 & 2 &&$-$0.05& $-$0.02 &    0.03 & 3 \\

[Co/Fe]        &~~0.20&   ~~0.19 &    0.06 & 6 &&~~0.11&  ~~0.11 &    0.05 & 4 \\

[Ni/Fe]        &~~0.02&   ~~0.01 &    0.08 & 15 &&~~0.01& $-$0.01 &    0.15 & 17 \\

[Ba/Fe]$_{\mathrm{II}}$ &~~0.17&   ~~0.19 &    0.06 & 3 &&~~0.29&  ~~0.34 &    0.09 & 3 \\

[La/Fe]$_{\mathrm{II}}$ &~~0.14&   ~~0.15 &    0.01 & 2 &&~~0.12&  ~~0.14 &    0.09 & 2 \\

[Eu/Fe]$_{\mathrm{II}}$ &~~0.32&   ~~0.31 &    0.04 & 2 &&~~0.29&  ~~0.30 & \nodata & 1 \\
\enddata
\end{deluxetable}

\clearpage

\begin{deluxetable}{lcccccccc}
\tabletypesize{\scriptsize}
\tablecaption{Mean elemental abundances of program clusters.}
\tablenum{9}
\tablewidth{0pc}
\tablehead{
\multicolumn{1}{c}{} &
\multicolumn{2}{c}{NGC~6287} &
\multicolumn{1}{c}{} &
\multicolumn{2}{c}{NGC~6293} &
\multicolumn{1}{c}{} &
\multicolumn{2}{c}{NGC~6541} \\
\cline{2-3}\cline{5-6}\cline{8-9}
\multicolumn{1}{c}{} &
\multicolumn{1}{c}{$\log g_{P}$} & \multicolumn{1}{c}{$\sigma$} &
\multicolumn{1}{c}{} &
\multicolumn{1}{c}{$\log g_{P}$} & \multicolumn{1}{c}{$\sigma$} &
\multicolumn{1}{c}{} &
\multicolumn{1}{c}{$\log g_{P}$} & \multicolumn{1}{c}{$\sigma$}}
\startdata
[Fe/H]$_{\mathrm{II}}$  & $-$2.01 & 0.06 && $-$1.99 & 0.00 && $-$1.76 & 0.04 \\

[O/Fe]  &  ~~0.31 & 0.06 && $-$0.14\tablenotemark{*} & \nodata && $-$0.33 & 0.25\\

[Na/Fe] &  ~~0.50 & 0.22 && ~~0.39 & 0.01 && ~~0.33 & 0.07 \\

[Mg/Fe] &  ~~0.35 & 0.09 && $-$0.04 & 0.22 && ~~0.34\tablenotemark{*} & \nodata \\

[Al/Fe] &  ~~0.93\tablenotemark{*} &  \nodata && ~~0.99 & 0.02 && ~~1.18 & 0.16 \\

[Si/Fe] &  ~~0.55 & 0.03 &&  ~~0.61 & 0.07 && ~~0.52 & 0.01 \\

[Ca/Fe] &  ~~0.21 & 0.04 &&  ~~0.23 & 0.01 && ~~0.34 & 0.04 \\

[Ti/Fe]$_{\mathrm{I}}$ & ~~0.11 & 0.09 && ~~0.10 & 0.02 && ~~0.20 & 0.03 \\

[Ti/Fe]$_{\mathrm{II}}$ & ~~0.14 & 0.03 && ~~0.12 & 0.03 && ~~0.22 & 0.05 \\

[V/Fe]  & $-$0.05 & 0.04 && ~~0.07 & 0.22 && $-$0.09 & 0.01 \\

[Cr/Fe] & $-$0.12 & 0.02 && $-$0.13 & 0.04 && ~~0.10 & 0.15 \\

[Co/Fe] &  ~~0.17 & 0.10 &&  ~~0.11 & 0.05 && ~~0.16 & 0.05 \\

[Ni/Fe] &  ~~0.04 & 0.06 && $-$0.03 & 0.01 && ~~0.02 & 0.01 \\

[Ba/Fe]$_{\mathrm{II}}$ & ~~0.39 & 0.05 && ~~0.00 & 0.07 && ~~0.23 &  0.06 \\

[La/Fe]$_{\mathrm{II}}$ & ~~0.28 & 0.02 && ~~0.15\tablenotemark{*} & \nodata && ~~0.13 & 0.01 \\

[Eu/Fe]$_{\mathrm{II}}$ & ~~0.45 & 0.01 && ~~0.42 & 0.01  && ~~0.31 & 0.02 \\
\enddata
\tablenotetext{*}{Single star measurement.}
\end{deluxetable}

\clearpage

\begin{deluxetable}{lccccccccccc}
\tabletypesize{\scriptsize}
\tablecaption{Abundance dependencies on model atmosphere.}
\tablenum{10}
\tablewidth{0pc}
\tablehead{
\multicolumn{1}{c}{} & \multicolumn{2}{c}{$\delta T_{eff}$} &
\multicolumn{1}{c}{} & \multicolumn{2}{c}{$\delta \log g$} &
\multicolumn{1}{c}{} & \multicolumn{2}{c}{$\delta v_{turb}$} &
\multicolumn{1}{c}{} & \multicolumn{2}{c}{$\delta T_{eff}/\delta \log g/\delta v_{turb}$}  \\
\cline{2-3}\cline{5-6}\cline{8-9}\cline{11-12}
\multicolumn{1}{c}{} & \multicolumn{1}{c}{$-$100} & \multicolumn{1}{c}{+100} &
\multicolumn{1}{c}{} & \multicolumn{1}{c}{$-$0.3} & \multicolumn{1}{c}{+0.3} &
\multicolumn{1}{c}{} & \multicolumn{1}{c}{$-$0.3} & \multicolumn{1}{c}{+0.3} &
\multicolumn{1}{c}{} & \multicolumn{1}{c}{$-$100} & \multicolumn{1}{c}{+100} \\
\multicolumn{1}{c}{} & \multicolumn{1}{c}{(K)} & \multicolumn{1}{c}{(K)} &
\multicolumn{1}{c}{} & \multicolumn{1}{c}{} & \multicolumn{1}{c}{} &
\multicolumn{1}{c}{} & \multicolumn{1}{c}{(km/s)} & \multicolumn{1}{c}{(km/s)}&
\multicolumn{1}{c}{} & \multicolumn{1}{c}{$-$0.1} & \multicolumn{1}{c}{+0.1} \\
\multicolumn{10}{c}{} & \multicolumn{1}{c}{$-$0.1} & \multicolumn{1}{c}{+0.1} }
\startdata
[Fe/H]$_{\mathrm{I}}$ &$-$0.11 & ~~0.13 && ~~0.00 & ~~0.01 && ~~0.07 & $-$0.06 && $-$0.09 & ~~0.10 \\

[Fe/H]$_{\mathrm{II}}$ &~~0.10 & $-$0.08 && $-$0.13 & ~~0.15 && ~~0.03 & $-$0.02 && ~~0.05 & $-$0.05 \\

 & & & & & & & & & & &\\

[O/Fe]         & $-$0.14 & ~~0.09 &&  ~~0.02 &$-$0.03 &&$-$0.04 & ~~0.01 && $-$0.13 & ~~0.10 \\

[Na/Fe]        &  ~~0.01 &$-$0.03 && ~~0.03 & $-$0.04 &&$-$0.07 & ~~0.05 && ~~0.00 & $-$0.02 \\

[Mg/Fe]        &  ~~0.03 &$-$0.06 && ~~0.01 & $-$0.03 &&$-$0.05 & ~~0.03 && ~~0.03 & $-$0.05 \\

[Al/Fe]        &  ~~0.01 &$-$0.03 && ~~0.03 & $-$0.03 &&$-$0.06 & ~~0.05 && ~~0.01 & $-$0.02 \\

[Si/Fe]        &  ~~0.13 &$-$0.13 && $-$0.04 & ~~0.04 &&$-$0.06 & ~~0.04 && ~~0.11 & $-$0.11 \\

[Ca/Fe]        & $-$0.03 & ~~0.01 && ~~0.02 & $-$0.04 && ~~0.05 & $-$0.05 && $-$0.01 & ~~0.01 \\

[Ti/Fe]$_{\mathrm{I}}$  & $-$0.09 & ~~0.08 && ~~0.03 & $-$0.03 &&$-$0.04 &~~0.04 && $-$0.09 & ~~0.08 \\

[Ti/Fe]$_{\mathrm{II}}$ &$-$0.06 &  ~~0.04 &&  ~~0.01 &$-$0.02 &&$-$0.02 &~~0.02 && $-$0.05 & ~~0.05 \\

[V/Fe]         & $-$0.11 & ~~0.10 && ~~0.03 & $-$0.02 &&$-$0.05 & ~~0.05 && $-$0.11 & ~~0.10 \\

[Cr/Fe]        & $-$0.03 & ~~0.01 && ~~0.02 & $-$0.03 &&$-$0.04 & ~~0.04 && $-$0.03 & ~~0.02 \\

[Co/Fe]        & $-$0.02 & ~~0.02 &&$-$0.01 & ~~0.02 &&$-$0.05 & ~~0.05 && $-$0.03 & ~~0.04 \\

[Ni/Fe]        &  ~~0.02 &$-$0.01 && $-$0.02 & ~~0.03 && ~~0.03 & $-$0.01 && ~~0.01 & $-$0.01 \\

[Ba/Fe]        & $-$0.13 & ~~0.12 &&  ~~0.04 &$-$0.06 && ~~0.25 & $-$0.23 && $-$0.07 & ~~0.03 \\

[La/Fe]        & $-$0.11 & ~~0.11 &&  ~~0.02 &$-$0.02 &&$-$ 0.01 & ~~0.02 && $-$0.10 & ~~0.10 \\

[Eu/Fe]        & $-$0.10 & ~~0.07 &&  ~~0.01 &$-$0.02 && $-$0.02 & ~~0.01 && $-$0.08 & ~~0.07 \\

 & & & & & & & & & & &\\

[Si~I/Ti~I]    & ~~0.23 & $-$0.21 && $-$0.06 & ~~0.06 && $-$0.02 & ~~0.01 && ~~0.20 & $-$0.18 \\

[Si~I/Ti~II]   & $-$0.01 & ~~0.03 && ~~0.09 & $-$0.08 && ~~0.00 & ~~0.00 && ~~0.01 & ~~0.00 \\

\enddata
\end{deluxetable}

\clearpage

\begin{deluxetable}{ccccccc}
\tabletypesize{\scriptsize}
\tablecaption{$\alpha$-element abundances}
\tablenum{11}
\tablewidth{0pc}

\tablehead{
\colhead{Id.} & 
\colhead{[Si/Fe]} &
\colhead{[Ca/Fe]} &
\colhead{[Ti/Fe]} &
\colhead{[$\alpha$/Fe]} &
\colhead{n\tablenotemark{1}}}
\startdata
NGC~6287 & 0.55 $\pm$ 0.03 & 0.21 $\pm$ 0.04 & 0.12 $\pm$ 0.07 & 
0.30 $\pm$ 0.02 & \\
NGC~6293 & 0.61 $\pm$ 0.07 & 0.23 $\pm$ 0.01 & 0.11 $\pm$ 0.03 & 
0.32 $\pm$ 0.02 & \\
NGC~6541 & 0.52 $\pm$ 0.01 & 0.34 $\pm$ 0.04 & 0.21 $\pm$ 0.05 & 
0.36 $\pm$ 0.02 & \\
&&&&&&\\
old halo   & 0.38 $\pm$ 0.05 & 0.30 $\pm$ 0.03 & 0.33 $\pm$ 0.03 & 
0.34 $\pm$ 0.02 & 15 \\
younger halo & 0.29 $\pm$ 0.09 & 0.14 $\pm$ 0.06 & 0.17 $\pm$ 0.07 & 
0.20 $\pm$ 0.05 & 7 \\
younger halo\tablenotemark{2} 
& 0.38 $\pm$ 0.09 & 0.21 $\pm$ 0.04 & 0.25 $\pm$ 0.06 & 
0.28 $\pm$ 0.02 & 5 \\
disk       & 0.39 $\pm$ 0.08 & 0.16 $\pm$ 0.07 & 0.31 $\pm$ 0.08 & 
0.29 $\pm$ 0.10 & 4 \\
\enddata
\tablenotetext{1}{Number of clusters.}
\tablenotetext{2}{Without Palomar~12 and Rupercht~106.}
\end{deluxetable}

\clearpage

\begin{figure}
\epsscale{1}
\figurenum{1}
\plotone{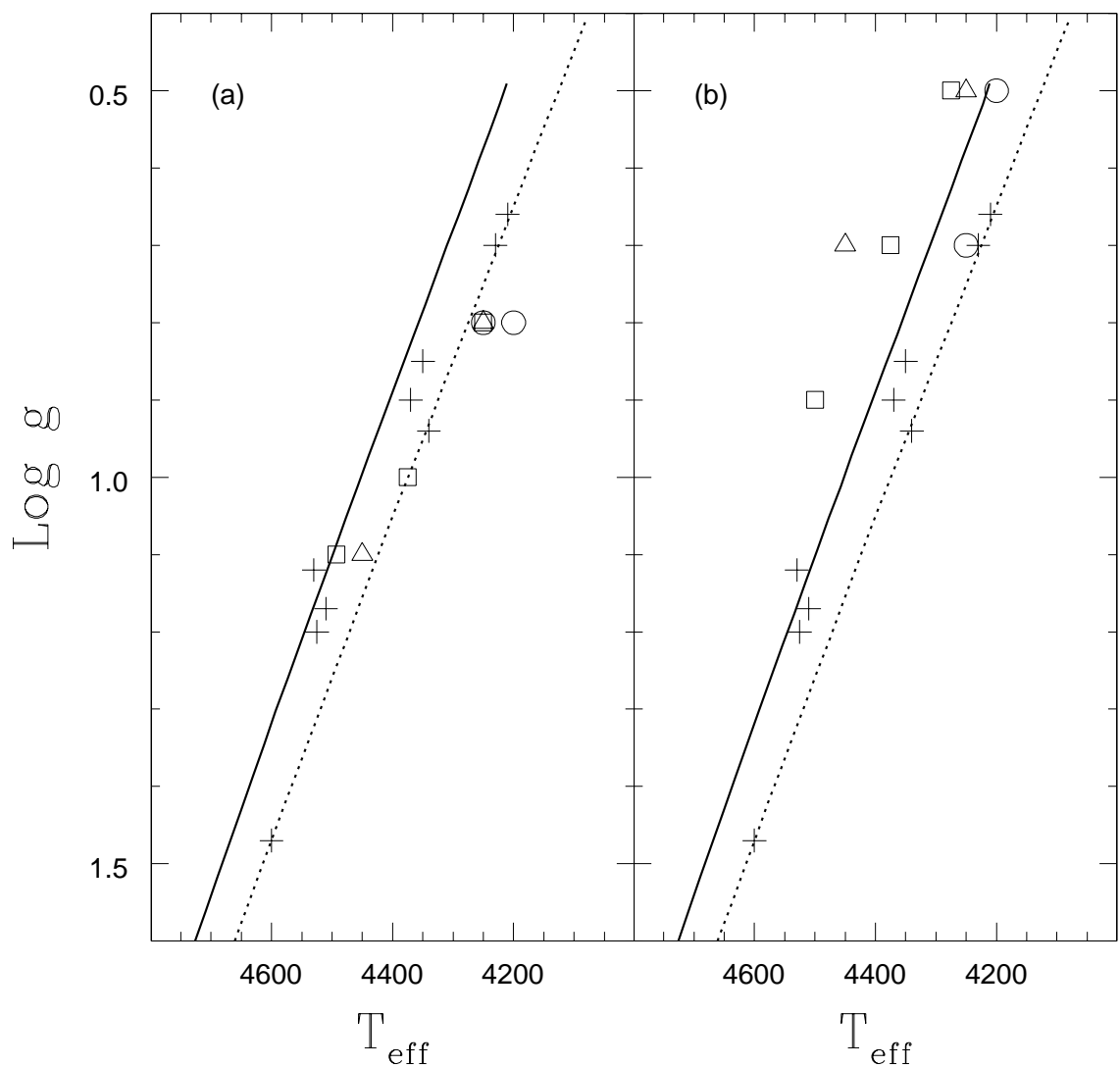}
\caption{Comparisons of the model atmosphere parameters to the model 
isochrones using photometric surface gravities (a) and spectroscopic 
surface gravities (b).
The open squares represent the NGC~6287 RGB stars, the open triangles
the NGC~6293 RGB stars, and the open circles the NGC~6541 RGB stars.
We also show the M92 RGB stars (Sneden et al. 1991) with crosses.
The solid lines is for the model isochrone with [Fe/H] = $-$2.14 and
the dotted lines for that with [Fe/H] = $-$1.84 
(Bergbusch \& VandenBerg 2001).}
\end{figure}

\begin{figure}
\epsscale{1}
\figurenum{2}
\plotone{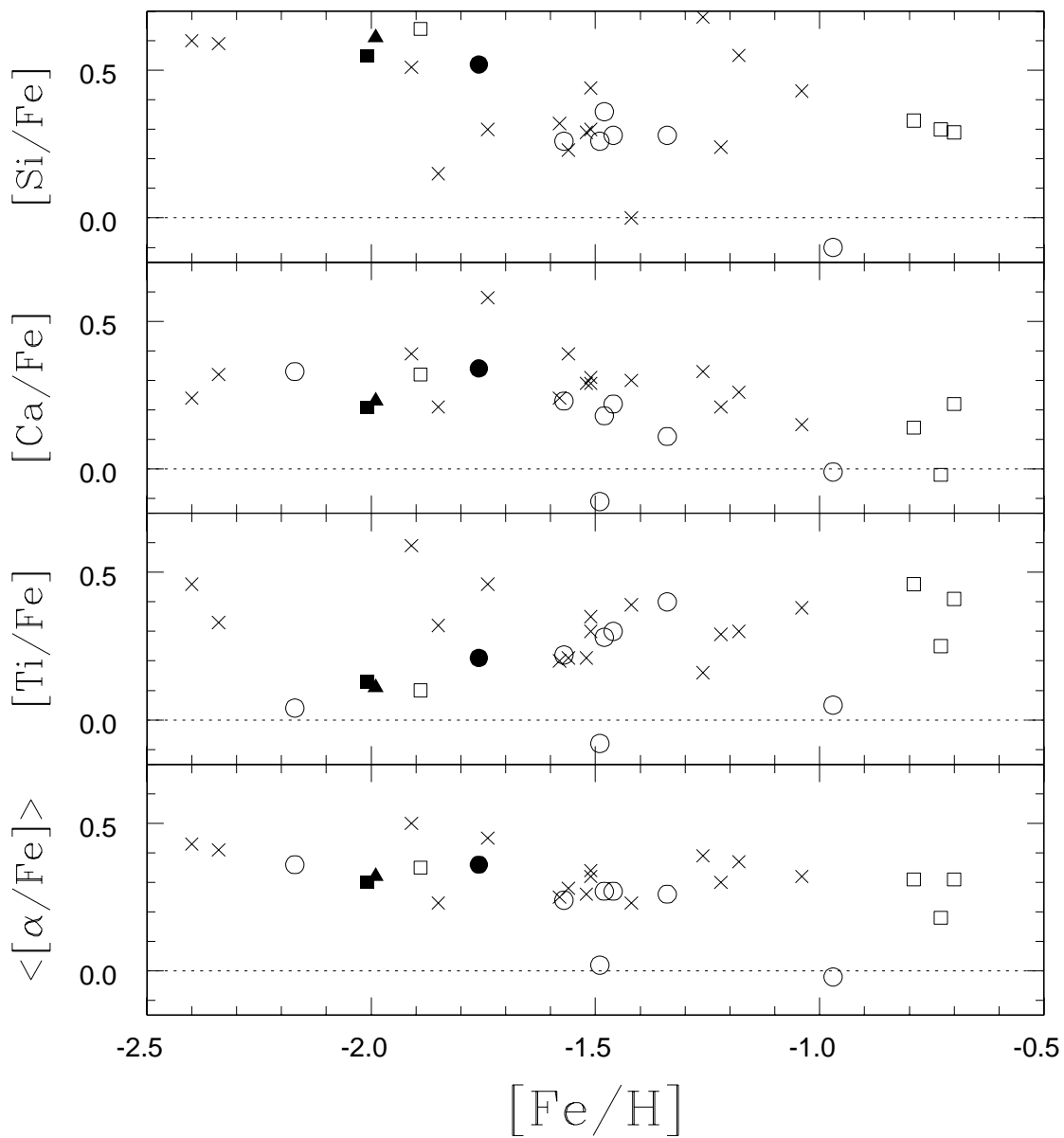}
\caption{$\alpha$-element abundances for globular clusters.
Crosses represent ``old halo" clusters, open circles 
``younger halo" clusters, and open squares ``thick disk" clusters.
NGC~6287 is represented by filled squares, NGC~6293 by filled triangles,
and NGC~6541 by filled circles. Our program clusters appear to be
silicon enhanced and titanium deficient.}
\end{figure}

\begin{figure}
\epsscale{1}
\figurenum{3}
\plotone{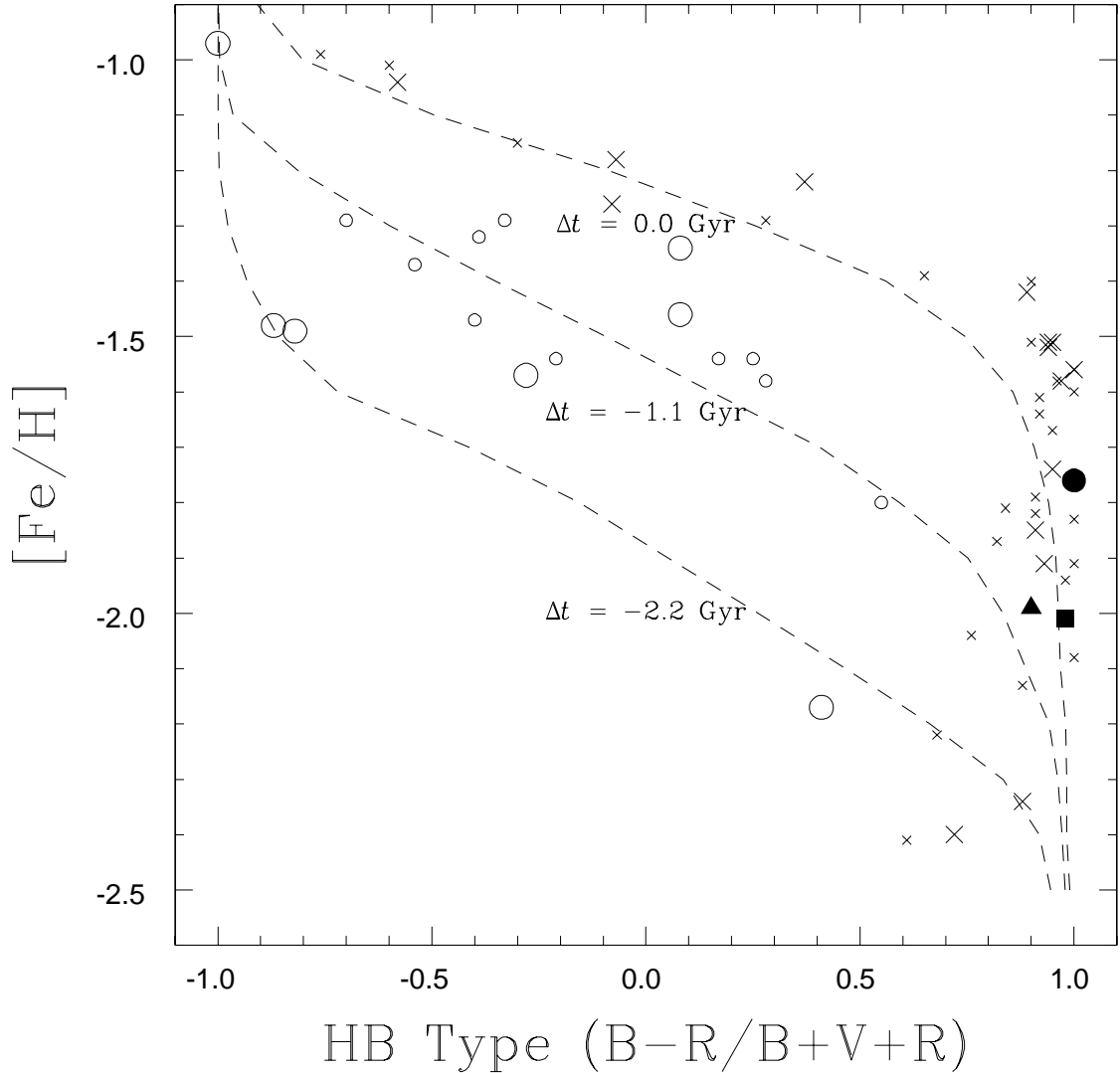}
\caption{HB type versus [Fe/H] (HB isochrones) of ``halo" globular clusters 
(Da Costa \& Armandroff 1995) and our program clusters.
Crosses are ``old halo" clusters and open circles are ``younger halo" clusters,
where large crosses and open circles denote clusters studied employing
high-resolution spectroscopy (see also Figure~2). 
NGC~6287 is represented by filled squares, 
NGC~6293 by filled triangles, and NGC~6541 by filled circles.
HB isochrones for $\Delta t$ = 0.0, $-$1.1, and $-$2.2 Gyr
(with respect to the mean age of ``old halo" globular clusters)
are also shown with dashed lines (Rey et al.\ 2001).}
\end{figure}

\begin{figure}
\epsscale{1}
\figurenum{4}
\plotone{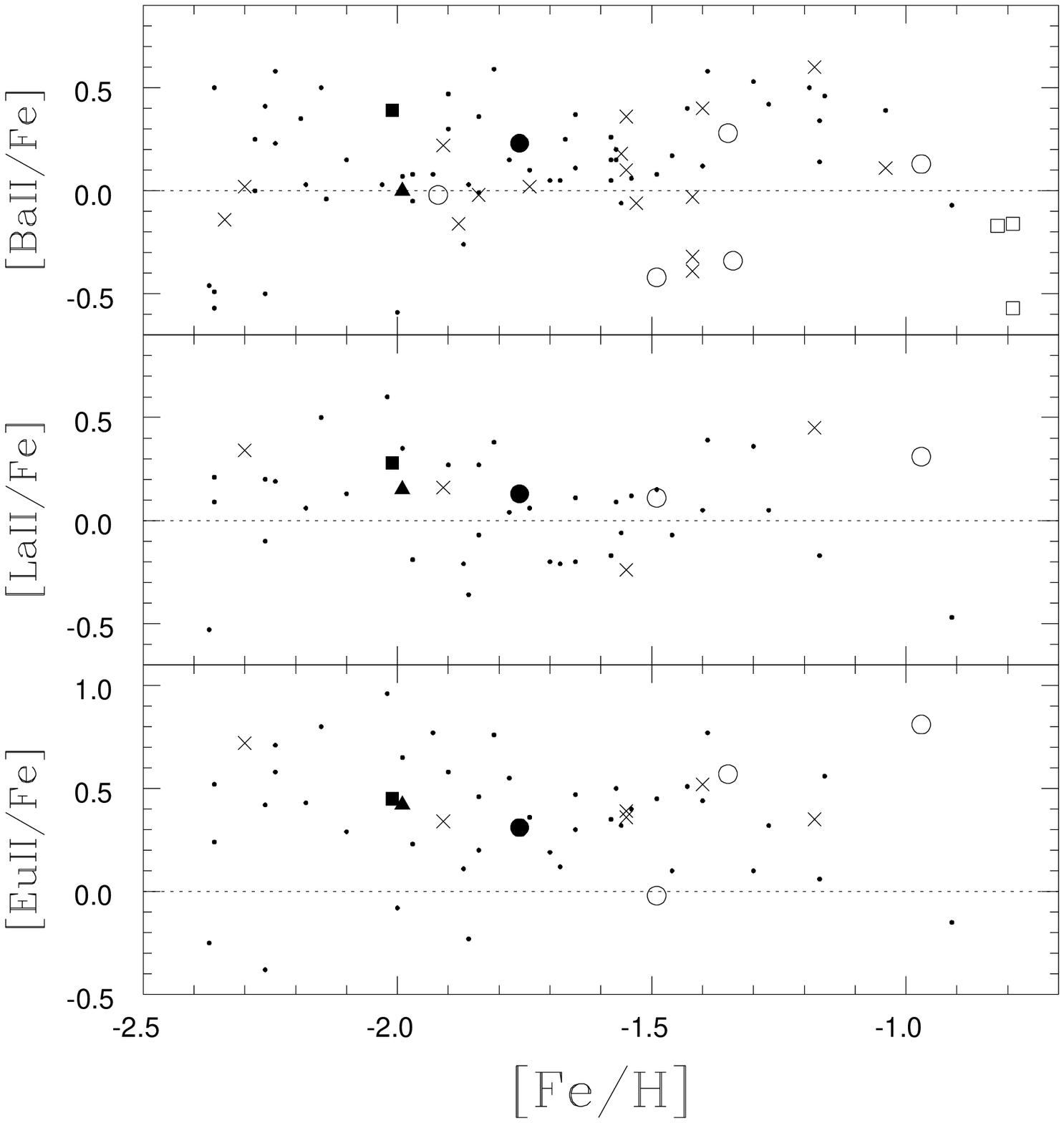}
\caption{Abundances of the neutron capture elements.
Crosses are ``old halo" clusters, open circles ``younger halo" clusters,
open squares ``thick disk" clusters, and dots the field stars
(Burris et al. 2000).
NGC~6287 is represented by filled squares, NGC~6293 by filled triangles,
and NGC~6541 by filled circles.}
\end{figure}

\begin{figure}
\epsscale{1}
\figurenum{5}
\plotone{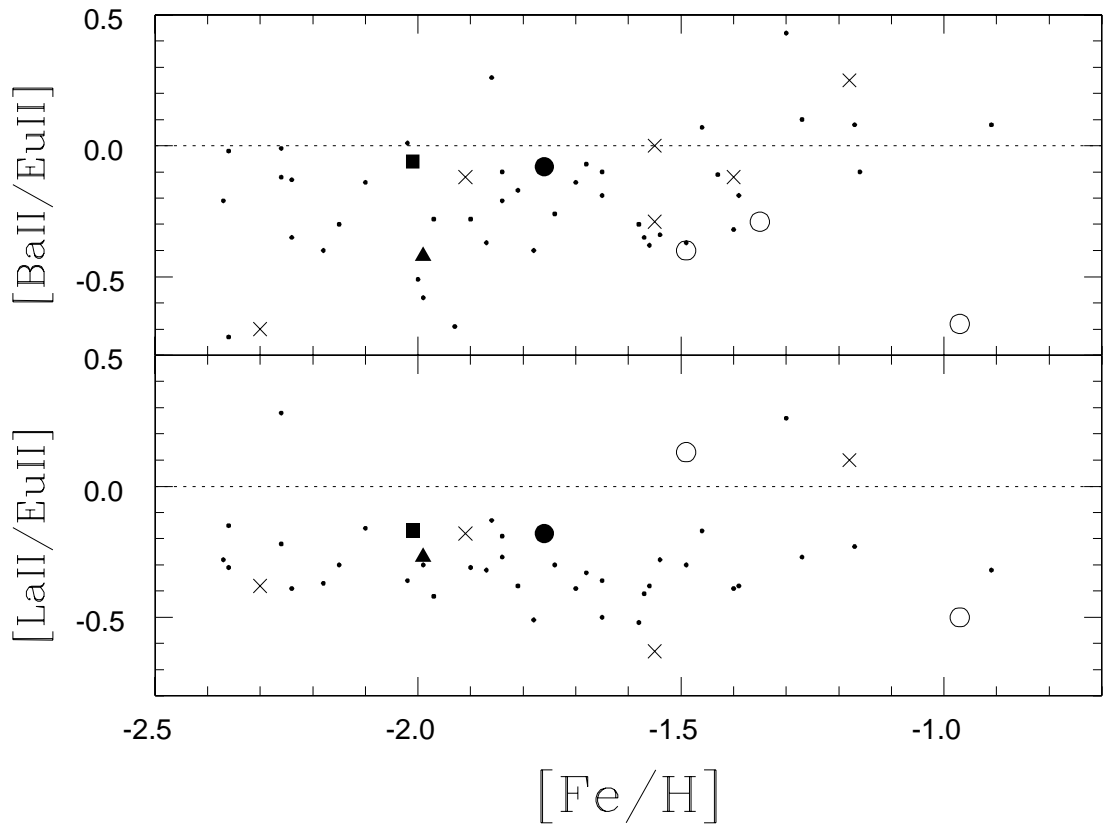}
\caption{Elemental ratios of the neutron capture elements.
Crosses are ``old halo" clusters, open circles ``younger halo" clusters,
open squares ``thick disk" clusters, and dots the field stars.
NGC~6287 is represented by filled squares, NGC~6293 by filled triangles,
and NGC~6541 by filled circles.}
\end{figure}

\begin{figure}
\epsscale{1}
\figurenum{6}
\plotone{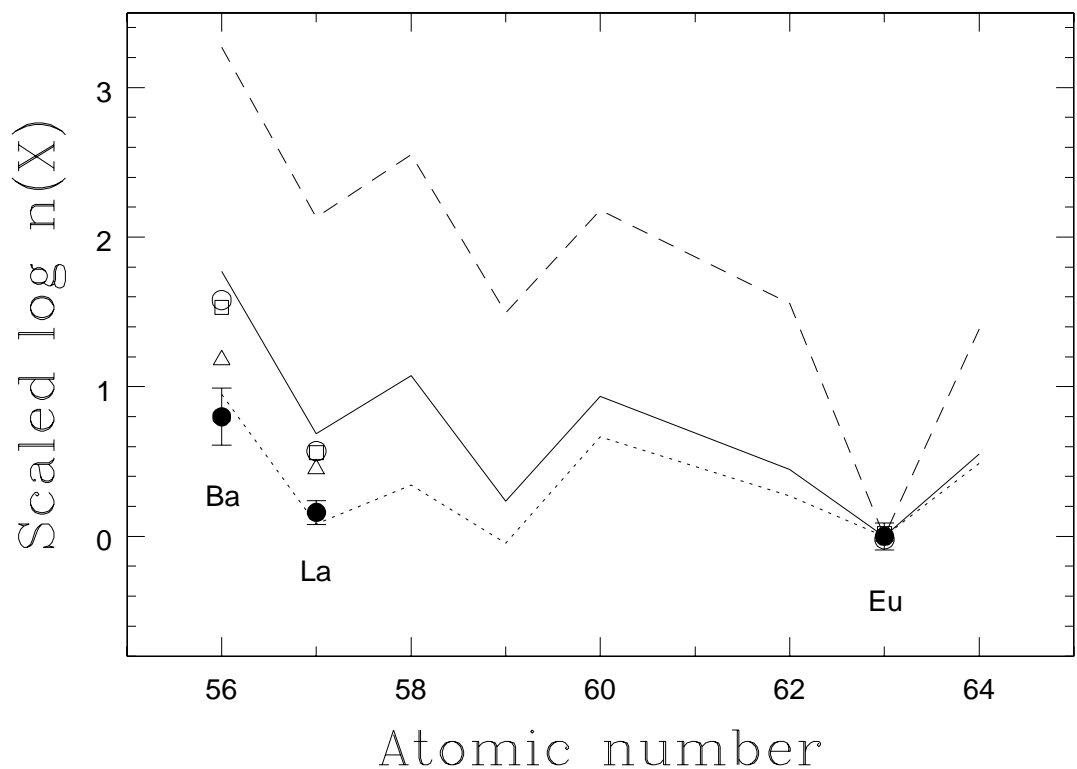}
\caption{Comparisons of the neutron capture elemental abundances
to those of the Sun.
Ba and La abundances are scaled to match Eu abundances.
Open squares are for NGC~6287, open triangles NGC~6293, 
and open circles NGC~6541.
M15 neutron capture elements are also plotted by filled circles 
(Sneden et al. 2000a).
The solid line represents the solar abundances,
the dotted line the solar $r$-process abundances,
and the dashed line the solar $s$-process abundances.}
\end{figure}

\begin{figure}
\epsscale{1}
\figurenum{7}
\plotone{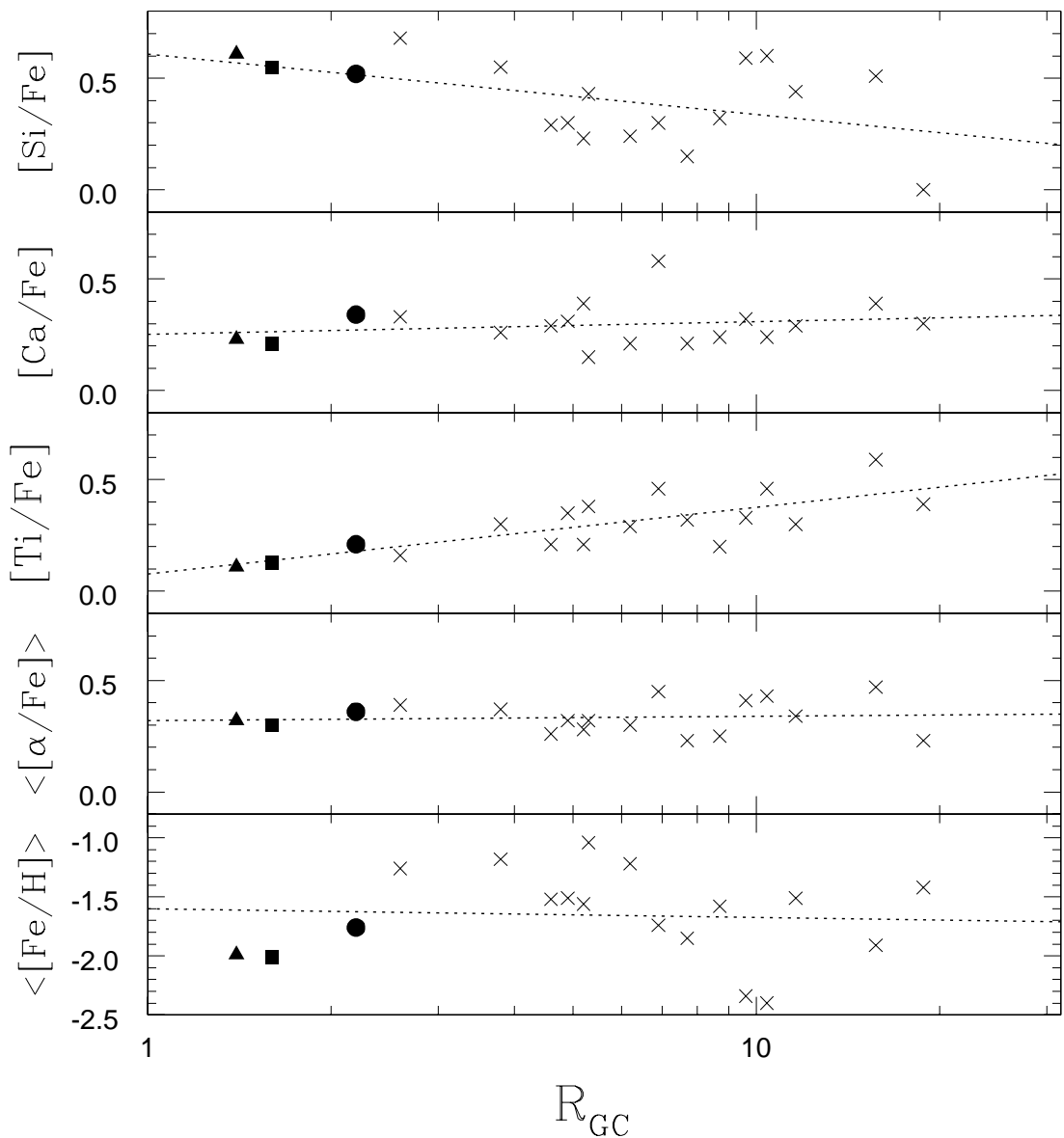}
\caption{The $\alpha$-element abundances and metallicities
of ``old halo" clusters as a function of $R_{GC}$.
NGC~6287 is represented by filled squares, NGC~6293 by filled triangles,
and NGC~6541 by filled circles.
The dotted lines represent the linear fit to the data.
Note that metallicities do not reflect the underlying
stellar population, but only the objects studied in our study.}
\end{figure}

\begin{figure}
\epsscale{1}
\figurenum{8}
\plotone{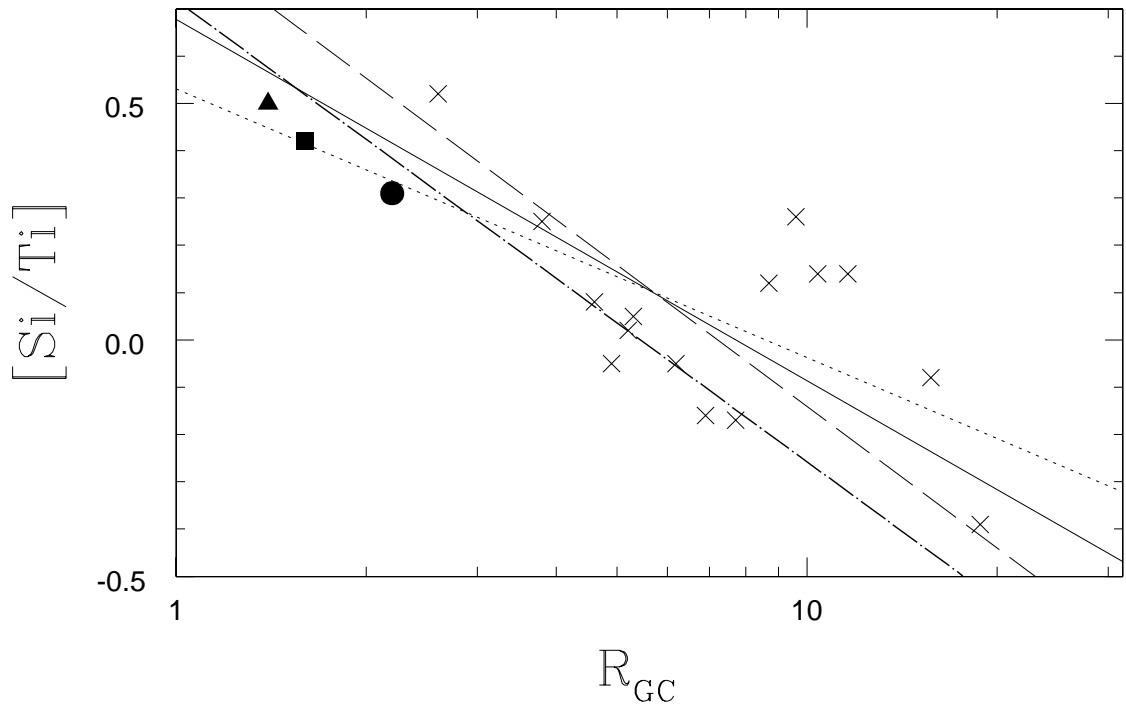}
\caption{[Si/Ti] as a function of $R_{GC}$.
Crosses represent ``old halo" clusters.
NGC~6287 is represented by filled squares, NGC~6293 by filled triangles,
and NGC~6541 by filled circles. The dotted line is for the linear fit,
the dashed line for the inverse linear fit, and the solid line for
the bisector linear fit to the data.
The dashed-dotted line represents the bisector linear fit to the clusters
with $R_{GC}$ $\leq$ 8 kpc (12 clusters).}
\end{figure}

\begin{figure}
\epsscale{1}
\figurenum{9}
\plotone{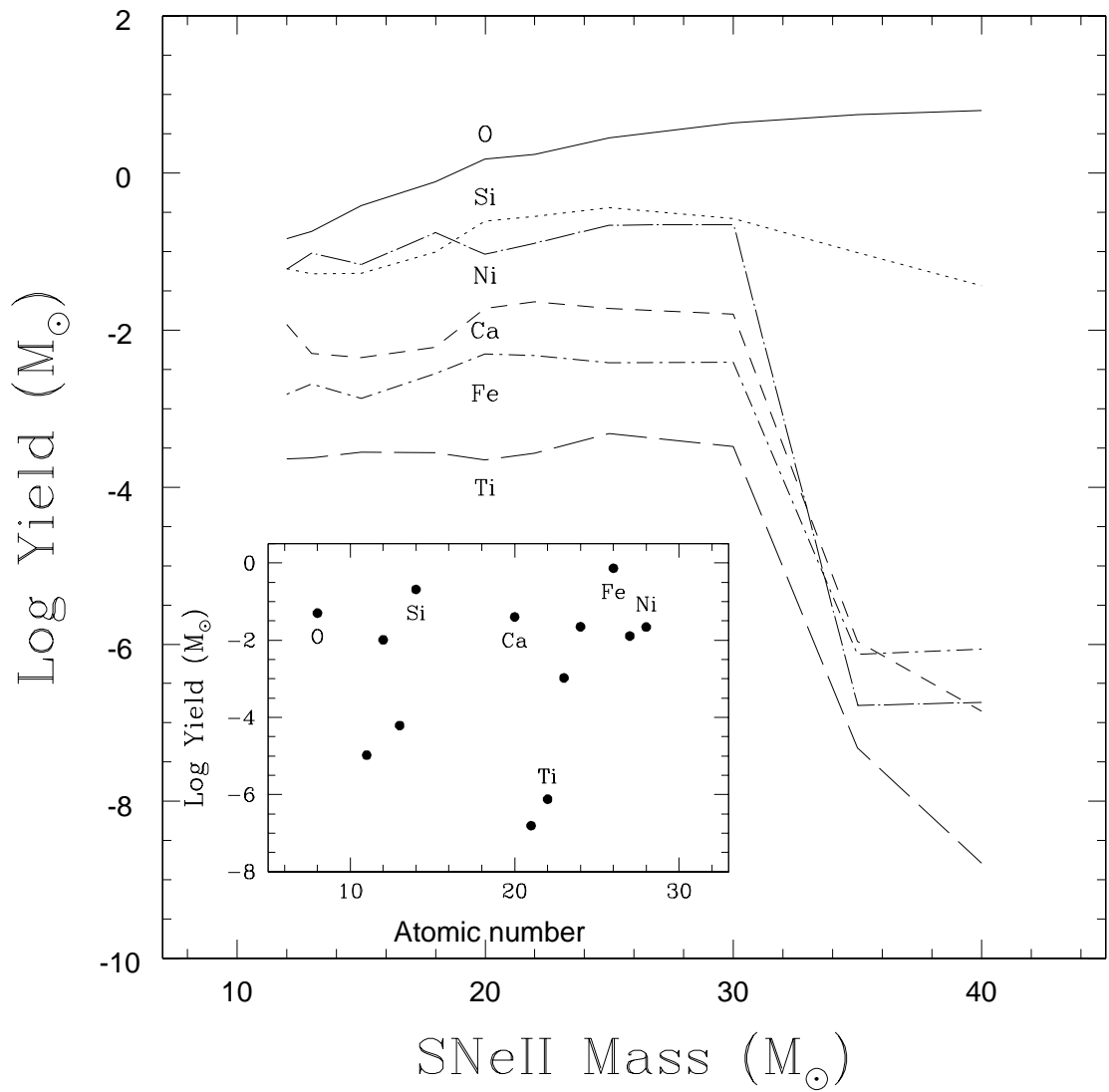}
\caption{SNe~II yields as a function of a progenitor mass
(Woosley \& Weaver 1995).
SNe~Ia yields as a function of atomic mass are also plotted in the inset.
For SNe~Ia, we adopt data from
H\"oeflich, Wheeler, \& Thielemann (1998).}
\end{figure}

\end{document}